\documentclass[journal=jacsat,manuscript=article, layout=traditional]{achemso}

\setkeys{acs}{maxauthors = 0}
\usepackage[version=3]{mhchem} 
\usepackage[LGRgreek]{mathastext}
\usepackage[percent]{overpic}
\usepackage{upgreek}
\usepackage{amsmath}

\author{Manasa Kaniselvan+}
\email{mkaniselvan@iis.ee.ethz.ch}
\affiliation[ETH Zürich]
{Integrated Systems Laboratory, Department of Information Technology and Electrical Engineering, ETH Zürich, CH-8092 Zürich, Switzerland}

\author{Kevin Portner+}
\email{kportner@iis.ee.ethz.ch}
\affiliation[ETH Zürich]
{Integrated Systems Laboratory, Department of Information Technology and Electrical Engineering, ETH Zürich, CH-8092 Zürich, Switzerland}

\author{Donato Francesco Falcone}
\email{dof@zurich.ibm.com}
\affiliation{IBM Research Europe - Zurich\unskip, 8803 Rüschlikon\unskip, Switzerland}

\author{Valeria Bragaglia}
\email{vbr@zurich.ibm.com}
\affiliation{IBM Research Europe - Zurich\unskip, 8803 Rüschlikon\unskip, Switzerland}

\author{Jente Clarysse}
\email{jente.clarysse@sl.ethz.ch}
\affiliation[ETH Zürich]
{Integrated Systems Laboratory, Department of Information Technology and Electrical Engineering, ETH Zürich, CH-8092 Zürich, Switzerland}

\author{Laura Bégon-Lours}
\email{lbegon@ethz.ch}
\affiliation[ETH Zürich]
{Integrated Systems Laboratory, Department of Information Technology and Electrical Engineering, ETH Zürich, CH-8092 Zürich, Switzerland}

\author{Marko Mladenović}
\email{mmladenovic@ethz.ch}
\affiliation[ETH Zürich]
{Integrated Systems Laboratory, Department of Information Technology and Electrical Engineering, ETH Zürich, CH-8092 Zürich, Switzerland}

\author{Bert J. Offrein}
\email{ofb@zurich.ibm.com}
\affiliation{IBM Research Europe - Zurich\unskip, 8803 Rüschlikon\unskip, Switzerland}

\author{Mathieu Luisier}
\email{mluisier@iis.ee.ethz.ch}
\affiliation[ETH Zürich]
{Integrated Systems Laboratory, Department of Information Technology and Electrical Engineering, ETH Zürich, CH-8092 Zürich, Switzerland}

\title[An \textsf{achemso} demo]
  { Electroforming Kinetics in HfOx/Ti RRAM: Mechanisms Behind Compositional and Thermal Engineering}

\begin{document}

\begin{abstract}

A critical issue affecting filamentary resistive random access memory (RRAM) cells is the requirement of high voltages during electroforming. Reducing the magnitude of these voltages is of significant interest, as it ensures compatibility with Complementary Metal-Oxide-Semiconductor (CMOS) technologies. Previous studies have identified that changing the initial stoichiometry of the switching layer and/or implementing thermal engineering approaches has an influence over the electroforming voltage magnitude, but the exact mechanisms remain unclear. Here, we develop an understanding of how these mechanisms work within a standard a-HfO$_x$/Ti RRAM stack through combining atomistic driven-Kinetic Monte Carlo (d-KMC) simulations with experimental data. By performing device-scale simulations at atomistic resolution, we can precisely model the movements of point defects under applied biases in structurally inhomogeneous materials, which allows us to not only capture finite-size effects but also to understand how conductive filaments grow under different electroforming conditions. Doing atomistic simulations at the device-level also enables us to link simulations of the mechanisms behind conductive filament formation with trends in experimental data with the same material stack. We identify a transition from primarily vertical to lateral ion movement dominating the filamentary growth process in sub-stoichiometric oxides, and differentiate the influence of global and local heating on the morphology of the formed filaments. These different filamentary structures have implications for the dynamic range exhibited by formed devices in subsequent SET/RESET operations. Overall, our results unify the complex ion dynamics in technologically relevant HfO$_x$/Ti-based stacks, and provide guidelines that can be leveraged when fabricating devices.
  
\end{abstract}


\newpage

\section{Introduction}

In filamentary resistive random access memory (RRAM), a few atomic relocations are sufficient to induce large conductance changes. This property allows the switching mechanism to be localized to the nanoscale \cite{Pi2018}, and the theoretical integration density of these memory devices to rival that of Flash \cite{Yoon2019}. Combining their scaling advantage with the ability to fabricate them from standard, Complementary Metal-Oxide-Semiconductor (CMOS)-compatible low-temperature-grown amorphous oxides, RRAM cells are well-suited for applications such as on-chip memory functionalities and neuromorphic computing primitives \cite{Christensen2022, Ielmini2018}.

However, the requirement of an electroforming process remains one of the most critical bottlenecks of the RRAM technology, preventing their adoption in commercial products. The operation of almost all filamentary RRAM devices begins with an initial electroforming step, where a relatively high voltage is applied to create a filament of conductive atomic defects that can be reversibly completed/disrupted during following SET and RESET operations. The morphology of the conductive filament(s) created during this process heavily influences all subsequent switching operations. Notably, the voltages required for this electroforming step are usually higher than those for SET and RESET processes \cite{Dittmann2021}, complicating the external circuitry required in practical applications. Due to finite-size effects, the magnitude of these voltages increases as devices are scaled down \cite{Chen2013}, or when the underlying switching material is more uniformly deposited, e.g., in case of a reduced density of grain boundaries \cite{An2024}. As the processes behind filament formation involve highly probabilistic movements of atoms through (typically) irregular lattices, the resulting conductive pathways are also formed differently in each device, leading to device-to-device variability. Approaching the technological readiness of RRAM devices will only become possible if the high electroforming voltages can be reduced and the initial filamentary growth process better controlled.

Experimentally, several methods have been shown to have a direct impact on the morphology of conductive filaments in RRAM cells. They can be roughly separated into compositional and thermal approaches. The former involves introducing structural components, either during growth or processing \cite{An2024, Xue2012}, to confine/guide \cite{Liu2010, Niu2016} or disperse \cite{Jeon2023} the formation of conductive filaments. On the other hand, thermal methods consist of adding heat trapping layers to the device \cite{Sarantopoulos2024, Wu2017}, sending pre-programming heating pulses \cite{Jeong2015}, or applying heat signals during operation \cite{Portner2021}. All of these mechanisms have been theorized to influence the way conductive filaments initially grow during electroforming, leading to differences in the subsequent High Resistance State (HRS)/ Low Resistance State (LRS) ratios, SET/RESET voltage magnitudes, or to reduced cycle-to-cycle variability. Nevertheless, their exact working principles are not yet well understood. 

Meanwhile, various computational modeling tools have been developed to better understand the operation of RRAM devices, involving different levels of simulation \cite{Sun2019, Urquiza2021, Kaniselvan2023}. However most of them rely on simplified compact models \cite{Woo2024}, continuum modeling approaches \cite{Zhang2022} or Kinetic Monte Carlo using uniform, three-dimensional (3D) grids, all of which involve abstracting away the underlying atomic structure. But since resistive switching behavior can be achieved within feature sizes as small as 2.1 $\times$ 2.1 nm\textsuperscript{2} \cite{Pi2018}, the motion of a few atoms can be sufficient to significantly alter the device resistance \cite{Aeschlimann2023}. The typical hourglass shape of these filaments \cite{Celano2014, Celano2015, Ma2020} further implies that the bottlenecks of their conductive pathways may be composed of only a few atomic defects. Because finite-size effects and the number of atoms participating in the electroforming of RRAM play an important role, it is necessary to describe them in simulation with an atomistic resolution. Only very few computational models are capable of operating at these scales. They generally suffer from high computational cost and are therefore restricted to small domain sizes \cite{Urquiza2021, Kaniselvan2023}.

Here, we use an atomistic driven-Kinetic Monte Carlo (d-KMC) solver to systematically explain the parameters determining the voltage required to electroform RRAM devices made of a standard, technologically-relevant amorphous hafnia (a-HfO$_x$) material stack. Such kinetic models are encoded with the mechanisms expected to affect resistive switching in RRAM cells at the level of individual atomic movements, and how they are influenced by externally applied fields. If all relevant events are present, experimental observables can be matched at the device-level. We use our d-KMC solver to reproduce trends in electroforming seen in measurements of fabricated devices based on the selected material stack. Specifically, we explore variation in the electroforming voltage (V$_{FORM}$) with device area, stoichiometry, sweep rate, and ambient temperature during the forming process. Together, these parameters provide a general overview of the different compositional and thermal tuning knobs that were previously used in the literature to modulate V$_{FORM}$ in HfO$_x$ RRAM devices. We thus present a realistic picture of conductive filament formation in a-HfO$_x$/Ti RRAM devices, considering the finite number and size of the atoms/ions involved in this process. 

\section{Results and Discussion}

\subsection{Correlating fabricated and simulated device domains}

\begin{figure}[t]
\centering\includegraphics[width=\textwidth]{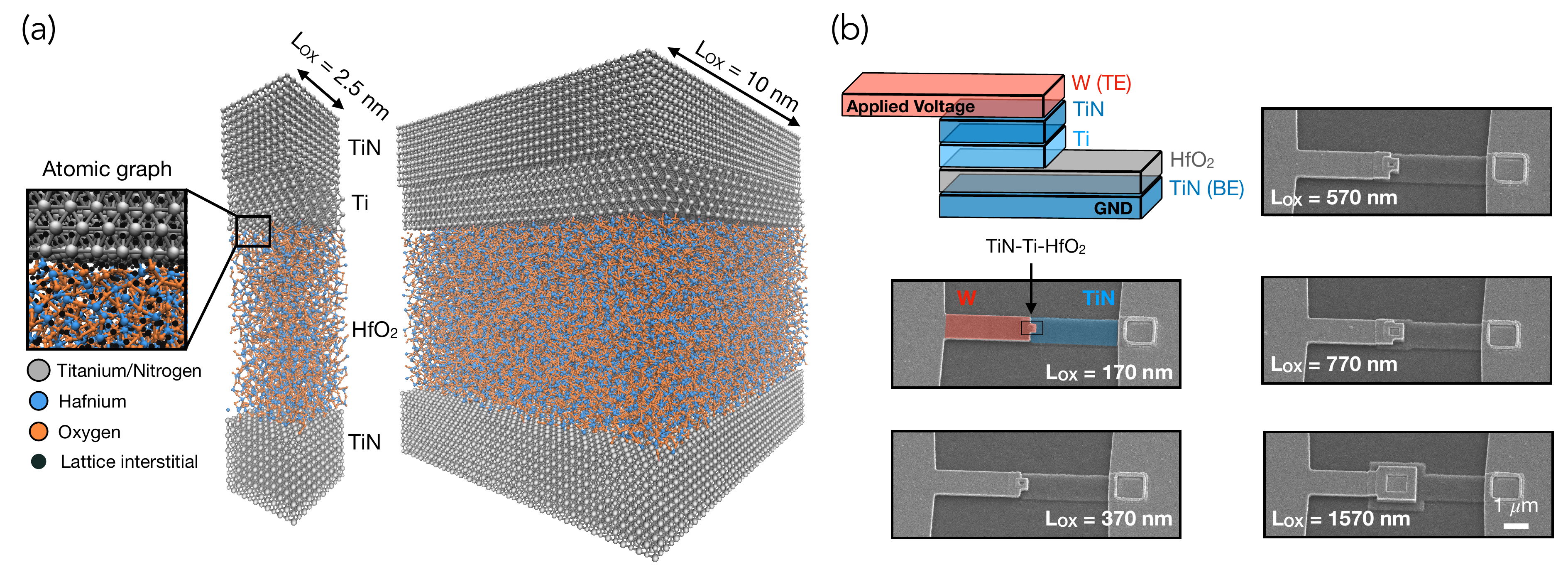}
        \caption{\textbf{Physical dimensions of the simulated and fabricated devices.} \textbf{(a)} Atomistically-resolved device stacks, with cross sections ranging from 2.5 $\times$ 2.5 nm\textsuperscript{2} to 10 $\times$ 10 nm\textsuperscript{2}. All atoms are color-coded. The zoom-in additionally shows interstitial positions at the a-HfO$_x$/Ti interface as black dots. \textbf{(b)} Schematic of the fabricated device stack, with the switching region corresponding to the device in (a), and scanning electron microscope (SEM) images providing an angled view of the fabricated devices. Their cross-sections range from 170 $\times$ 170 nm\textsuperscript{2} to 1.57 $\times$ 1.57 \textmu m\textsuperscript{2}. The scale bar for all images is 1 \textmu m.}
        \label{fig:devices}
\end{figure}

The mechanisms determining conductive filament formation strongly depend on the exact device size, material stack, structural irregularity of the switching layer, and interface configuration. To bridge experimentally-observed trends with the underlying kinetics of atomic rearrangement through atomistic simulations, we therefore first need to achieve a correspondence between the fabricated and modeled devices. The device stack considered is TiN/a-HfO$_x$/Ti/TiN (a-HfO$_x$ represents the amorphous-phase hafnia), where the Ti layer functions as a reservoir to collect migrating oxygen ions. A schematic of this stack is shown in \textbf{Fig. \ref{fig:devices}(a)}. The density of the simulated a-HfO$_x$ is 10.7 g/cm$^3$, matching that of the deposited material, as measured by X-Ray Reflectivity (XRR) (Supporting Information, \textbf{Fig. S1}).

While the thickness of the a-HfO$_x$ switching region can be reached by d-KMC simulations, the lateral device cross section cannot be exactly reproduced due to either the computational cost of modeling areas larger than 10 $\times$ 10nm\textsuperscript{2} at atomistic resolution or the difficulty of fabricating ultrascaled RRAM devices. The lateral dimensions, however, directly affect the magnitude of V$_{FORM}$ \cite{Chen2013}. We therefore analyzed the characteristics of fabricated devices (see \textbf{Fig. \ref{fig:devices}(b)}) with areas ranging from 1.57 $\times$ 1.57 \textmu m\textsuperscript{2} down to 170 $\times$ 170 nm\textsuperscript{2}, all made of the same material stack. Note that the active switching area is the region between the overlap of the bottom and top electrodes, and is laterally isolated to ensure that the conductive filaments are confined within this area. If measurements at different sizes follow a clear trend, these trends can be applied to the ultrascaled dimensions that are achievable via d-KMC simulations, linking the experimental and simulated trends.

\subsubsection{Fabricated devices}

We fabricate TiN/a-HfO$_x$/Ti/TiN/W stacks with CMOS and back-end-of-line (BEOL) compatible processes, where the tungsten (W) capping layer prevents oxidation of the TiN electrode \cite{Falcone2023}. Further details about the fabrication are provided in the Methods section, and in Refs. \cite{Stecconi2024, Stecconi2022}. Our devices are operated by grounding the TiN bottom electrode (BE) and applying an electrical signal to the W-TiN top electrode (TE). A conductive filament is initially created during an electroforming step by applying a positive V$_{FORM}$ to the W-TiN TE. To explore the effect of the a-HfO$_x$ stoichiometry, we consider two different vacancy/oxygen concentrations, which we refer to as `standard' and `leaky'. The latter is grown in a reduced-O$_2$ environment (1 s of O$_2$ plasma time during Plasma-Enhanced Atomic Layer Deposition (PEALD) for the leaky stack instead of 10 s for the standard one). The leaky HfO$_x$ exhibits a reduced atomic density caused by a lower oxygen concentration and thus a higher density of oxygen vacancies, as indicated by X-Ray Reflectivity (XRR) measurements (\textbf{Fig. S1} of the Supporting Information).

\subsubsection{Modeling approach}

All RRAM simulations begin with an atom-by-atom construction of the material stack. The amorphous-phase of the a-HfO$_x$ switching layer is obtained using melt-quench processes in Molecular Dynamics, as detailed in our previous work \cite{Kaniselvan2023}. Interstitials defects within the device atomic structure are subsequently identified through a Voronoi approach, which defines positions that diffusing oxygen atoms can occupy. To track structural changes during resistive switching, we use an atomistic d-KMC simulation framework capable of modeling the full operating range of RRAM devices, including electroforming, SET/RESET \cite{Kaniselvan2023}, and pulsed conductance modulation \cite{Kaniselvan2024}. The underlying KMC code implementation has been optimized to treat up to a million atoms \cite{kaniselvan2024accelerated}, enabling an entirely atomistic simulation of domain sizes up to 5 $\times$ 10 $\times$ 10 nm\textsuperscript{3}. The ability to treat these scales allows us to match record-low sizes of HfO$_2$ RRAM fabricated using advanced techniques. For example: \textbf{Ref \citenum{Pi2018}} (4 $\times$ 2.1 $\times$ 2.2 nm\textsuperscript{3}), \textbf{Ref \citenum{Govoreanu2011}} (10 $\times$ 10 $\times$ 5 nm\textsuperscript{3}), and \textbf{Ref \citenum{KaiShinLi2014}} ( 6 $\times$ 4 $\times$ 5 nm\textsuperscript{3} to 6 $\times$ 1 $\times$ 3 nm\textsuperscript{3}). The application has also been extended to include the effects of Joule heating in the bulk oxide (details are provided in the Methods section). 

Under an applied voltage, the a-HfO$_x$/Ti switching region undergoes a series of discrete structural changes. Each of such events involves ions and/or vacancies being generated, diffusing, or recombining based on activation energies computed with the Nudged Elastic Band (NEB) method \cite{neb}, and modified by local potential and heat distributions. To include all known relevant effects, we incorporate the reduced activation energy of ion-vacancy generation at the Ti interface \cite{OHara2014}, and the increased energy for ion diffusion within the oxidized Ti layer (Supporting Information, \textbf{Table S1}). Every structural change is followed by a re-computation of the local electrostatic potential, heat, and electrical current. An overview of the full simulation approach is provided in \textbf{Fig. S2(a)} of the Supporting Information. The parameters of the internal current solver, which considers tunneling through the potential barriers induced by the applied field, are calibrated against quantum transport calculations on a stoichiometric a-HfO$_x$ structure (Supporting Information, \textbf{Fig. S2(b)}). More information about the domain construction and simulation approach is in the Methods section.

\subsection{Kinetics of electroforming}

\begin{figure}[t]
\centering\includegraphics[width=\textwidth]{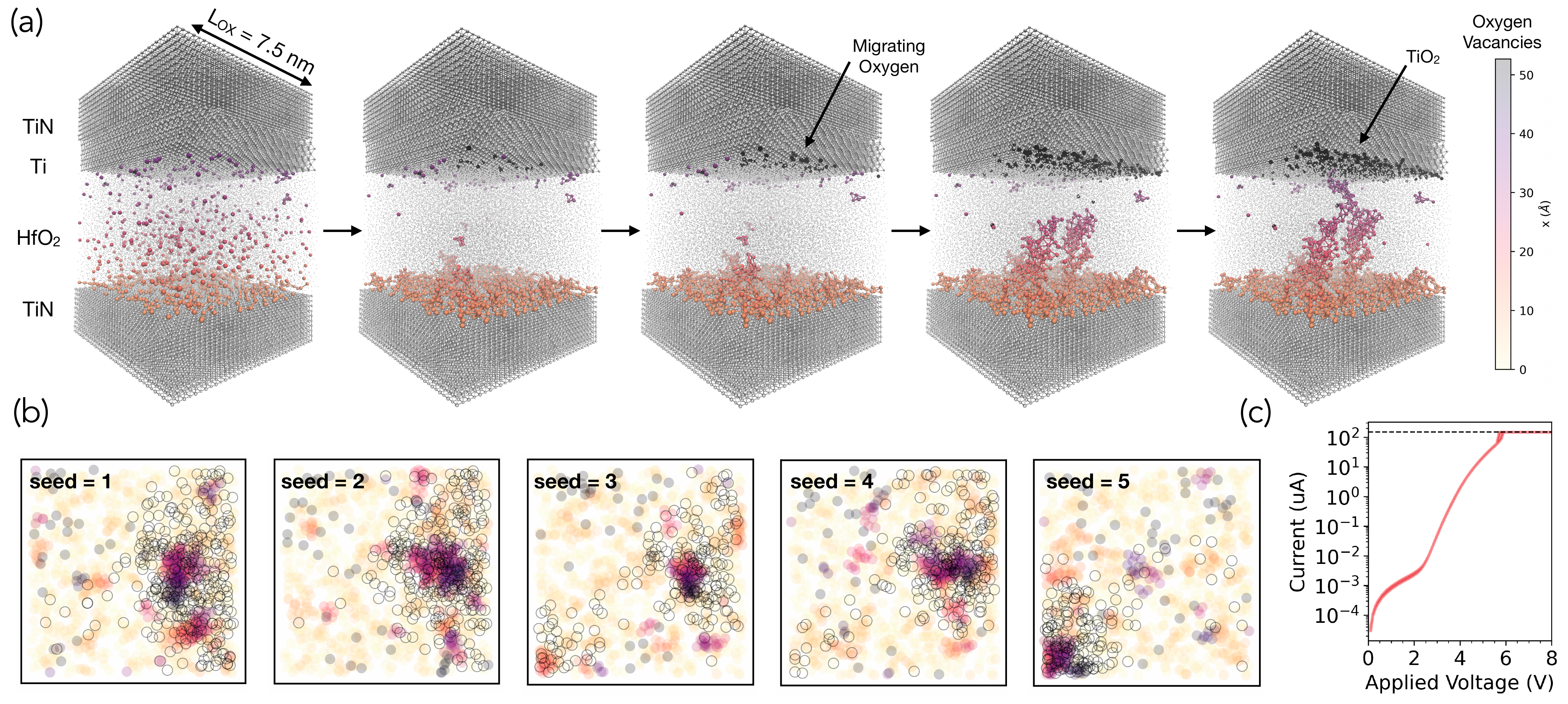}
        \caption{\textbf{Modeling the influence of structural inhomogeneity on conductive filament formation in TiN/a-HfO$_x$/Ti/TiN RRAM devices.} \textbf{(a)} Snapshots extracted along the electroforming process of a device with an area of 7.5 $\times$ 7.5 nm\textsuperscript{2}. The colored spheres indicate the locations of oxygen vacancies (V$_o$) and oxygen in the Ti reservoir (TiO$_x$), whereas the grey ones refer to Ti and N atoms in the contacts. The Hf and O atoms are pictured with points in order to more clearly visualize the distribution of vacancies inside the device. The locations of V$_o$ are colored according to their depth along the $x$-axis connecting the top and bottom electrodes. \textbf{(b)} Top-view plots of five electroforming processes performed on the same device, but with different random seeds. The first one corresponds to the last snapshot in (a). The colored spheres indicate V$_o$, following the same color scheme as in (a). Oxygen atoms which have migrated from their original lattice positions are pictured as black circles. \textbf{(c)} Current-voltage (I-V) characteristic of the the same devices as in (b). The spread is caused by the random seeds. The compliance current I$_{CC}$ is set to 150 \textmu A.}
        \label{fig:forming_timeline}
\end{figure}

When electroforming the a-HfO$_x$/Ti stack, oxygen is extracted from the a-HfO$_x$ into an oxidizable reservoir, leaving behind a conductive filament of oxygen vacancies (V$_o$) in the bulk oxide. Previous work has gathered extensive experimental evidence for this oxygen-extraction process \cite{Bertaud2012} and imaged the filamentary nature of the resulting conductive regions \cite{Privitera2013-vc}. These imaging studies also show that the conductive filaments tend to have an hourglass shape \cite{Celano2014, Celano2015, Ma2020} and typically break at one end, opening a tunneling gap that impedes current flow \cite{Hubbard2021}. Oxygen can then be re-incorporated from one interface back into the lattice to close this gap. The extraction and re-incorporation process is facilitated in RRAM device stacks by incorporating oxidizable reservoirs (typically Ti, Hf) at one interface that can trap and store migrating oxygen ions \cite{Dittmann2021, Fang2014}. They also serve to further locally reduce the energy required to generate V$_o$ \cite{OHara2014}, where dangling bonds and high potential fluctuations are present \cite{OHara2014}.

In \textbf{Fig. \ref{fig:forming_timeline}(a)} we show snapshots along an electroforming process for a simulated TiN/a-HfO$_x$/Ti/TiN RRAM device with an area of 7.5 $\times$ 7.5 nm\textsuperscript{2}. At voltages much below V$_{FORM}$, the energy is insufficient to generate new V$_o$, but existing charged V$_o$ drift under the applied field and either form immobile clusters \cite{Kamiya2013, Schmitt2017}, or pile up at the TiN cathode where they nucleate the bases of potential conductive filaments. As the voltage increases, the first few oxygen vacancies are generated at the HfO$_x$/Ti interface, because of the reduced activation energy to pull oxygen atoms out of the surface HfO$_x$. These oxygen atoms can then oxidize the neighboring Ti reservoir. Although they are relatively mobile in the bulk a-HfO$_x$, they become more static once they form a TiO$_x$ layer. This prevents them from directly recombining with the charged V$_o$ left behind, which `drift' to the counter TiN electrode under the influence of the applied field through cascaded exchanges of their positions with neighborhood oxygen atoms. The V$_o$ are reduced at the TiN cathode, and the potential of these early conductive pathways align with that of the cathode potential, creating lateral potential variations and preferentially accelerating the growth of existing filaments \cite{Dittmann2021}.

The observed kinetic mechanism behind the formation of conductive filaments in RRAM cells is consistent with Molecular Dynamics (MD) simulations of switching in small device stacks \cite{Urquiza2021}, models of oxygen diffusion through vacancies in substoichiometric a-HfO$_x$ \cite{Clima2012}, and temporally-resolved imaging studies of fabricated HfO$_x$ devices \cite{Hubbard2021}. We note, however, that the microscopic switching mechanisms seem to be specific to the materials involved in each RRAM, and cannot be straightforwardly extrapolated from one stack to the other. For example, the migration of metal cations, while shown to be negligible in a-HfO$_x$ via MD simulations \cite{Xiao2019} and Electron Beam-Induced Current (EBIC) imaging \cite{Hubbard2021}, may play a role in other materials such as TaO$_x$ \cite{TsurumakiFukuchi2023, Ma2020}. 

\subsubsection{Capturing structural irregularity}

To demonstrate that our simulations capture the effect of local structural irregularity inherent to amorphous switching materials, in \textbf{Fig. \ref{fig:forming_timeline}(b)-(c)} we show five identical electroforming processes on the same TiN/a-HfO$_x$/Ti/TiN stack (with area 7.5 $\times$ 7.5 nm\textsuperscript{2} and t$_{ox}$ = 5 nm), but with different internal random seeds. The oxide switching layer has an initial stoichiometry of a-HfO$_{x=1.9}$, corresponding to an oxygen vacancy concentration of 5\% \cite{Govoreanu2011}. The random seeds determine the initial distribution of native V$_o$ in the oxide, as well as the Monte Carlo-based selection of random numbers during the structure perturbation step. We do not introduce any seeding processes to guide conductive filament formation in specific locations.

Despite the different initial V$_o$ distributions and stochastic nature of the Monte Carlo processes, a filament tends to grow in the same location of this structure, indicating the existence of preferential atomic diffusion pathways within the underlying atomic graph. Specifically, the combination of realistically distributed local densities, bond lengths, and oxygen coordination numbers (Supporting Information, \textbf{Fig. S2(c)}), introduces structural irregularities in the simulation domain that bias filament growth towards specific regions. Filaments also typically form near the contact edges due to the higher field localization in these areas, which is consistent with previous reports of forming depending on the perimeter of the electrode in nanoscale RRAM \cite{Nauenheim2010}. Importantly, our model allows for the growth of multiple partially complete filaments if the simulation domain is made large enough ($>$ 5 $\times$ 5 nm$^2$). Such phenomena have been imaged experimentally in HfO$_2$ RRAM \cite{Privitera2013-vc}. 

\subsection{Oxide stoichiometry}

\begin{figure}[H]
\centering\includegraphics[width=\textwidth]{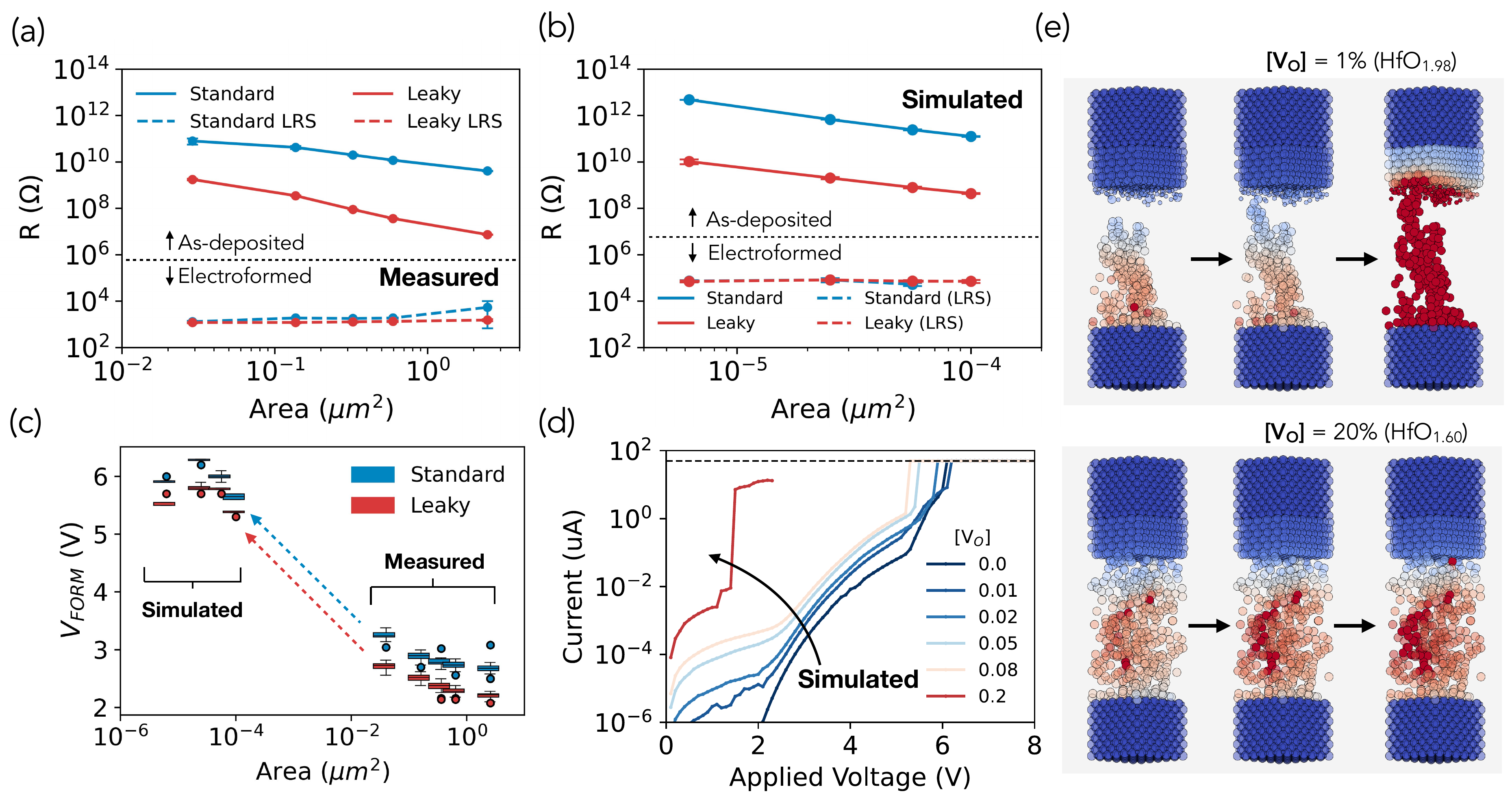}
        \caption{\textbf{Influence of device area and oxide stoichiometry on the electroforming process.} \textbf{(a)} Current measured from the as-deposited (solid lines) and electroformed (dashed lines) fabricated devices, for the standard (blue lines) and leaky (red lines) oxide samples as a function of the device area. \textbf{(b)} Simulated results under the same conditions as in (a), for devices with areas of 2.5 $\times$ 2.5 nm\textsuperscript{2} to 10 $\times$ 10 nm\textsuperscript{2}. In (a)-(b) the error bars indicate the median and standard deviation across (a) 36 different devices and (b) 5 different random seeds. \textbf{(c)} Measured V$_{FORM}$ for the devices in (a) and (b), extracted at the point at which a compliance current I$_{CC}$ = 20 \textmu A is reached. Both the measured and simulated data are shown on the same plot, separated by area range. The measured data corresponds to the same 36 devices as in (a). \textbf{(d)} Simulated forming I-V sweeps for devices with area 2.5 $\times$ 2.5 nm\textsuperscript{2} with different initial V$_o$ concentration from 0 (HfO$_2$) to 20\% (HfO$_{1.6}$). \textbf{(e)} Snapshots of the atomic structure of the device in (d) taken along the electroforming process with (top) [V$_o$]=1\% and (bottom) [V$_o$]=20\%. The V$_o$, migrating oxygen atoms, and contact atoms are colored according to their instantaneous effective temperature induced during the forming process. Hf and lattice-O atoms are omitted for clarity.}
        \label{fig:stoichiometry}
\end{figure}

Next, we examine the electrical properties of the standard and leaky a-HfO$_x$ devices in \textbf{Fig. \ref{fig:stoichiometry}(a)-(b)}, and compare the measured trends with those of smaller-area simulated devices using two different oxide stoichiometries, HfO$_2$ (``standard'') and HfO$_{1.9}$ (``leaky''). The latter is achieved by removing 5\% of the oxygen atoms from oxide layer. The resistance of the fabricated devices with pristine, as-deposited a-HfO$_x$ layers is measured under a bias of 2 V, which remains below the range of V$_{FORM}$. The induced currents are, however, high enough that the influence of measurement noise becomes negligible. The LRS resistance after electroforming is extracted at a readout voltage of V$_{read}$ = 0.2 V. 

The larger number of oxygen vacancies in the leaky oxide results in the pristine current of the corresponding stack to be $\sim$100$\times$ higher than that of the devices made of standard oxides. The pristine resistance of both standard and leaky oxides is also proportional to their area, as expected. Meanwhile, in the LRS, the resistance of devices with areas between 170 $\times$ 170 nm\textsuperscript{2} and 1.57 $\times$ 1.57 \textmu m\textsuperscript{2} remains almost constant, and the values from the standard and leaky oxides cannot be distinguished. This area-independence points to a filamentary electroforming mechanism, after which current flows primarily through a highly localized conductive pathway. The area dependence (independence) of the pristine (electroformed) state is reproduced in simulations, albeit at different resistance magnitudes. We attribute this difference in part to the chosen current compliance (I$_{CC}$), which also determines the exact magnitude of the LRS resistance. This cannot be exactly standardized between the simulated and experimentally measured devices, as in the former the compliance is imposed instantaneously while in the latter there are typically finite delays that affect the LRS. 

To transition them into their LRS, the devices in \textbf{Fig. \ref{fig:stoichiometry}(a)} undergo an electroforming process. Different biasing schemes have been proposed to do this \cite{Stathopoulos2019}, but here we opt for a Ramped Voltage Stress (RVS) scheme where a constant voltage ramp of 0.2 V/s is applied across the contacts to induce soft dielectric breakdown within the oxide. Each electroforming process is performed under an I$_{CC}$ of 20 \textmu A, with a 10 $k\Omega$ external series resistor. Further details on the measurement setup are provided in the Methods section. The V$_{FORM}$ of the leaky oxides are noticeably lower than those of the standard oxide (\textbf{Fig. \ref{fig:stoichiometry}(c)}), as observed in other works on breakdown in a-HfO$_x$ \cite{Zhao2024, Sharath2014-sg, Sharath2014}. 

After electroforming, all fabricated devices made of both standard and leaky oxides can be reliably cycled between their ON (LRS) and OFF (HRS) states in subsequent SET/RESET measurements.  Notably, the SET voltages (V$_{SET}$) do not show an area dependence (Supporting Information, \textbf{Fig. S3}), indicating that the length of the induced tunneling gap between the tip of the formed filament on the counter electrode may be similar in all devices. However, we found that the leaky devices present a slightly lower dynamic range (HRS/LRS ratio) (Supporting Information \textbf{Fig. S4(a)}). 

The transition from area-dependent (pristine) to area-independent (LRS) current results from the shift from trap-assisted tunneling between well-separated point defects (e.g., V$_o$) to nearly Ohmic conduction through filaments. The tendency of devices with reduced areas to require higher electroforming voltages V$_{FORM}$ has been observed across multiple works \cite{AnChen}. This ``finite size'' effect has been attributed to the lower likelihood of smaller structures to contain highly defective areas capable of nucleating conductive filaments. As a result, electroforming voltages of 2 to 3 V in devices at the micron scale translate into $>$ 5 V when their area is scaled down to the nanometer regime \cite{Pi2018, Govoreanu2011}. These finite size effects are least pronounced in uniformly amorphous compounds such as the ones considered here, and enhanced in polycrystalline materials due to selective electroforming at grain boundary locations \cite{Winkler2022, Petzold2019}. Deviation from this trend occurs in the smallest (simulated) devices with area 2.5 $\times$ 2.5 nm\textsuperscript{2} (\textbf{Fig. \ref{fig:stoichiometry}(c)}). At these scales, the surface area of lateral boundaries becomes significant compared to the volume of the switching material, leading to filaments preferably forming at edges/corners of the stack where potential variations can be more sharply confined. As a result, the forming voltages decrease once again. Additionally, V$_{FORM}$ levels off in very large area devices, as observed in measurements. These larger domains sample the full range of different coordination environments within the switching layer.

The fact that reduced oxygen (increased $V_o$) concentration leads to lower forming voltages and HRS/LRS ratios has been previously observed \cite{Jana2024}. Going one step further, we can investigate through simulation the different electroforming kinetics with respect to stoichiometric changes to the initial oxide. In \textbf{Fig. \ref{fig:stoichiometry}(d)} we plot RVS forming curves using a sweep rate of 0.2 V/s for structures with initial V$_o$ concentration between 0\% and 20\%. We use the device with the smallest area of 2.5 $\times$ 2.5 nm\textsuperscript{2} to better visualize the change in V$_o$ arrangements. The corresponding atomic structures taken at different time-points along the electroforming process are shown in \textbf{Fig. \ref{fig:stoichiometry}(e)}. Each atom is colored according to its instantaneous effective temperature, which indicates where current flow is spatially concentrated. Note that this temperature, resulting from the solution of the Fourier Heat equation with the power dissipated by current flow at each point, is only `effective' because temperature is a macroscopic quantity which cannot be defined at the atomic level. It is highest where hot atoms are ``bottlenecked'' and thus traces the locations of conductive filaments. 

In the devices with more stoichiometric oxides, the growth of conductive filaments primarily involves the vertical drift of vacancies (via substitution of nearby oxygen atoms) across the oxide. During this process, the first layer of the Ti electrode oxidizes into TiO$_x$. Meanwhile in the highly substoichiometric oxide, native V$_o$ in the oxide can form percolating pathways across shorter distances, thus inducing lateral fields and further encouraging lateral migration of V$_o$. Hence, the voltage required for electroforming is no longer dominated by the capacity to generate new V$_o$, as in standard HfO$_2$, but by the energy required to migrate existing V$_o$ and turn the primarily vertical direction of propagation of point defects (along the direction of the external field) to a lateral movement. At 20\% V$_o$, the device eventually exhibits a ``forming-free'' behavior where subsequent SET/RESET processes occur within the same voltage range as required for electroforming. Since there is minimal oxygen extraction from the oxide into a reservoir layer, filaments formed in this way can only undergo thermally-induced RESET operations. This dependence of filament growth kinetics on stoichiometry may consolidate the primarily vertical filamentary growth observed in many studies \cite{Hubbard2021, Kaniselvan2023} with the primarily lateral dynamics observed in others; for example, in the SET process of the HfO$_x$/Hf device in \textbf{Ref \citenum{Urquiza2021}}, and in the similar stack of \textbf{Ref \citenum{Kumar2016}} where radial oxygen migration was observed.

\subsection{RVS sweep rate}

\begin{figure}[H]
\centering\includegraphics[width=\textwidth]{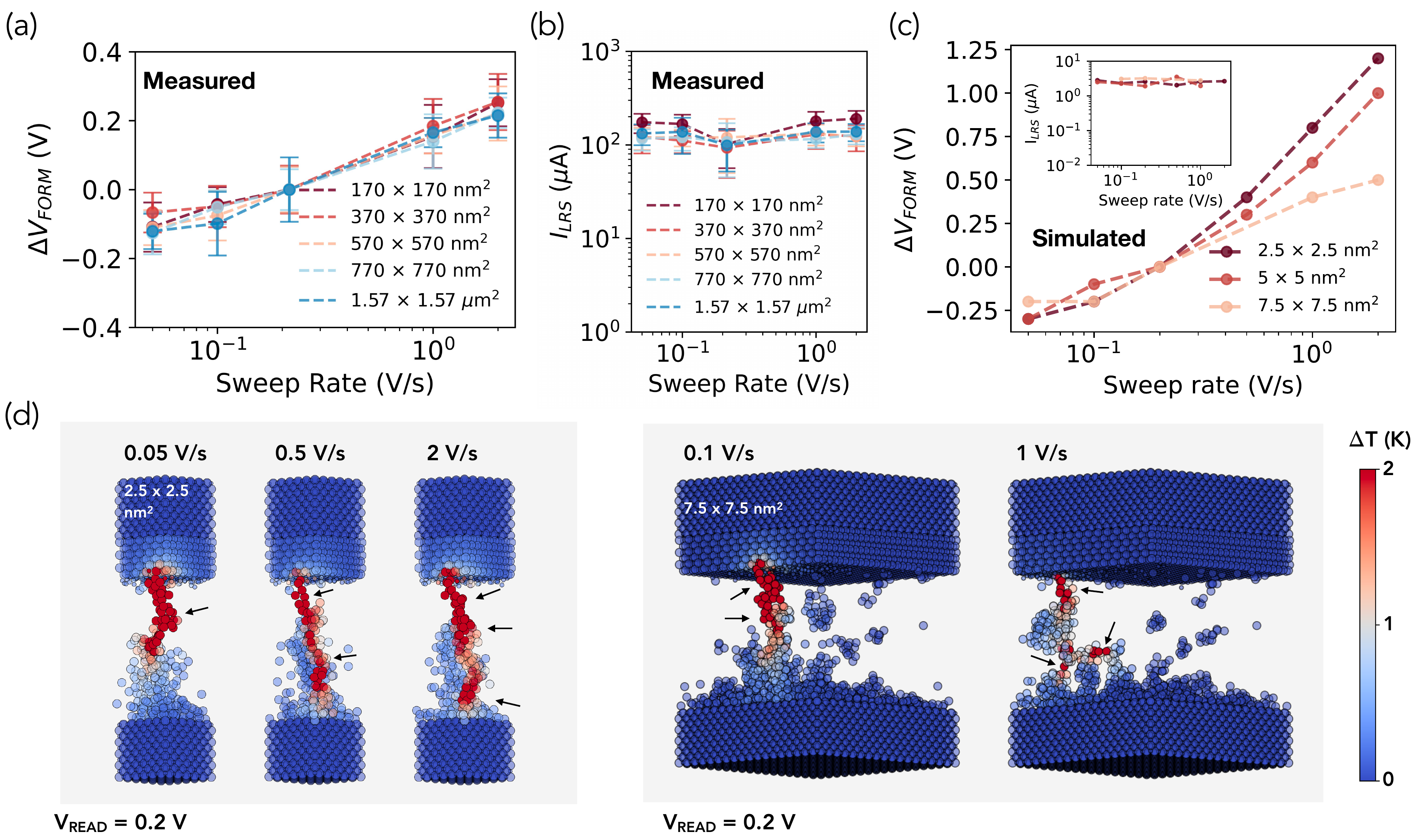}
        \caption{\textbf{Trade-offs between voltage and sweep rates}. \textbf{(a)} Change in measured V$_{FORM}$ for the standard TiN/a-HfO$_x$/Ti/TiN stack at different RVS sweep rates, compared to a reference rate of 0.2 V/s. \textbf{(b)} LRS currents extracted from the devices in (a) at V$_{read}$ = 0.2 V. Errors bars represent the standard deviation across 27 different samples. \textbf{(c)} Simulated variation in forming voltage for devices with areas of 2.5 $\times$ 2.5 nm\textsuperscript{2} to 7.5 $\times$ 7.5 nm\textsuperscript{2}. Results are compared to a reference rate of 0.2 V/s. For each device, the same random seed is used in all simulations, and only the time per voltage point is varied. The inset shows the LRS currents of the formed devices at V$_{read}$ = 0.2 V. \textbf{(d)} Example vacancy distributions in the structures formed at different sweep rates from the data in (c), for areas of  2.5 $\times$ 2.5 nm\textsuperscript{2} and 7.5 $\times$ 7.5 nm\textsuperscript{2}. The V$_o$, migrating oxygen atoms, and contact atoms are colored according to their effective temperature induced by V$_{read}$. Hf and lattice-O atoms are omitted for clarity.}
        \label{fig:sweeprate}
\end{figure}

The magnitude of the voltage triggering the atomic rearrangements behind electroforming depends on the time during which this voltage is applied, due to a known voltage-time trade-off \cite{RodriguezFernandez2017, Lorenzi2015, Luo2012}. In typical RVS forming procedures, V$_{FORM}$ can be controlled by the voltage sweep rate. In \textbf{Figs. \ref{fig:forming_timeline}-\ref{fig:stoichiometry}} we used a fixed sweep rate of 0.2 V/s for both the fabricated and simulated devices. In \textbf{Fig. \ref{fig:sweeprate}} we now investigate the influence of different sweep rates on the resulting V$_{FORM}$ and on the nature of the formed conductive filaments. Experimentally, faster sweep rates result in electroforming occurring at higher voltages (\textbf{Fig. \ref{fig:sweeprate}(a)}), without substantially changing the conductance of the final LRS (\textbf{Fig. \ref{fig:sweeprate}(b)}) under the chosen I$_{CC}$. At very slow sweep rates (i.e., 0.05 V/s), V$_{FORM}$ approaches a minimum value. In simulations, we explore a range of sweep rates for devices with area up to 7.5 $\times$ 7.5 nm\textsuperscript{2} (\textbf{Fig. \ref{fig:sweeprate}(c)}). We use a fixed voltage ramp of 0.1 V per step, and vary the duration for which the device is held at each voltage (e.g., 0.05 s per 0.1 V for a sweep rate of 2 V/s, or 2 s per 0.1 V for a sweep rate of 0.05 V/s). To better isolate the impact of the sweep rate on the variation in V$_{FORM}$ and the LRS filament morphology, the same random seed is used in all cases. 

The simulations follow the experimental trend, with higher forming voltages being required at higher sweep rates and vice versa. Similarly to the experimental results, a minimum V$_{FORM}$ is reached at low sweep rates, below which the device does not have sufficient energy to overcome the energy barrier for the creation of point defects. On the other hand, at high sweep rates, the duration at which the device is held at each voltage is too short to drive all of the required atomic generation and migration processes at the voltages where they are first able to occur. The oxide instead encounters much higher voltages which induce significant Joule heating. Under these conditions, V$_o$ undergo both directional drift and thermally-driven diffusion. The impact of Joule heating is less significant for devices with smaller areas, as their currents in the pristine state, before filament formation, are lower. As a consequence, smaller devices can withstand higher voltages before thermally-driven processes kick in, leading to a more pronounced influence of the sweep rate on $\Delta$V$_{FORM}$. At even higher sweep rates than shown in \textbf{Fig. \ref{fig:sweeprate}(c)} (e.g., $>$ 2 V/s), we start seeing generation of defect pairs in the bulk oxide, where local temperatures are highest, which eventually leads to hard dielectric breakdown of the device \cite{Padovani2024}.

Visualizing the distribution of current flow in the LRS provides further insight into the nature of conductive filaments formed at different sweep rates. In \textbf{Fig. \ref{fig:sweeprate}(d)} we plot the induced local temperature distributions under an LRS readout operation (V$_{read}$ = 0.2 V) for structures formed at different sweep rates. The simulated temperatures under V$_{read}$ remain relatively low, up to 310 K within the filament, and, as mentioned before, serve as an indirect trace of current flow pathways. In all cases, the current flow is bottle-necked near the reservoir layer, where the highest electric fields and thus the most highly-directional movement of vacancies occurs during electroforming. However, in the devices formed under faster sweep rates (i.e., higher V$_{FORM}$), high-temperature points are also visible throughout the oxide layer, indicating that current flow is laterally-concentrated through narrow pathways of point defects. In contrast, the flow of current in the devices formed under slower sweep rates (i.e., lower V$_{FORM}$) is instead distributed across wider pathways of V$_o$. These results are consistent with the picture that a lower V$_{FORM}$ induces a ``stronger'' single conductive filament, while higher V$_{FORM}$ tends to create multiple or weaker conductive pathways. Note that the ultimate value of the LRS current still remains the same in all cases (inset of \textbf{Fig. \ref{fig:sweeprate}(c)}), since it is determined primarily by the chosen I$_{CC}$.

\subsection{Electroforming temperature}

\begin{figure}[H]
\centering\includegraphics[width=\textwidth]{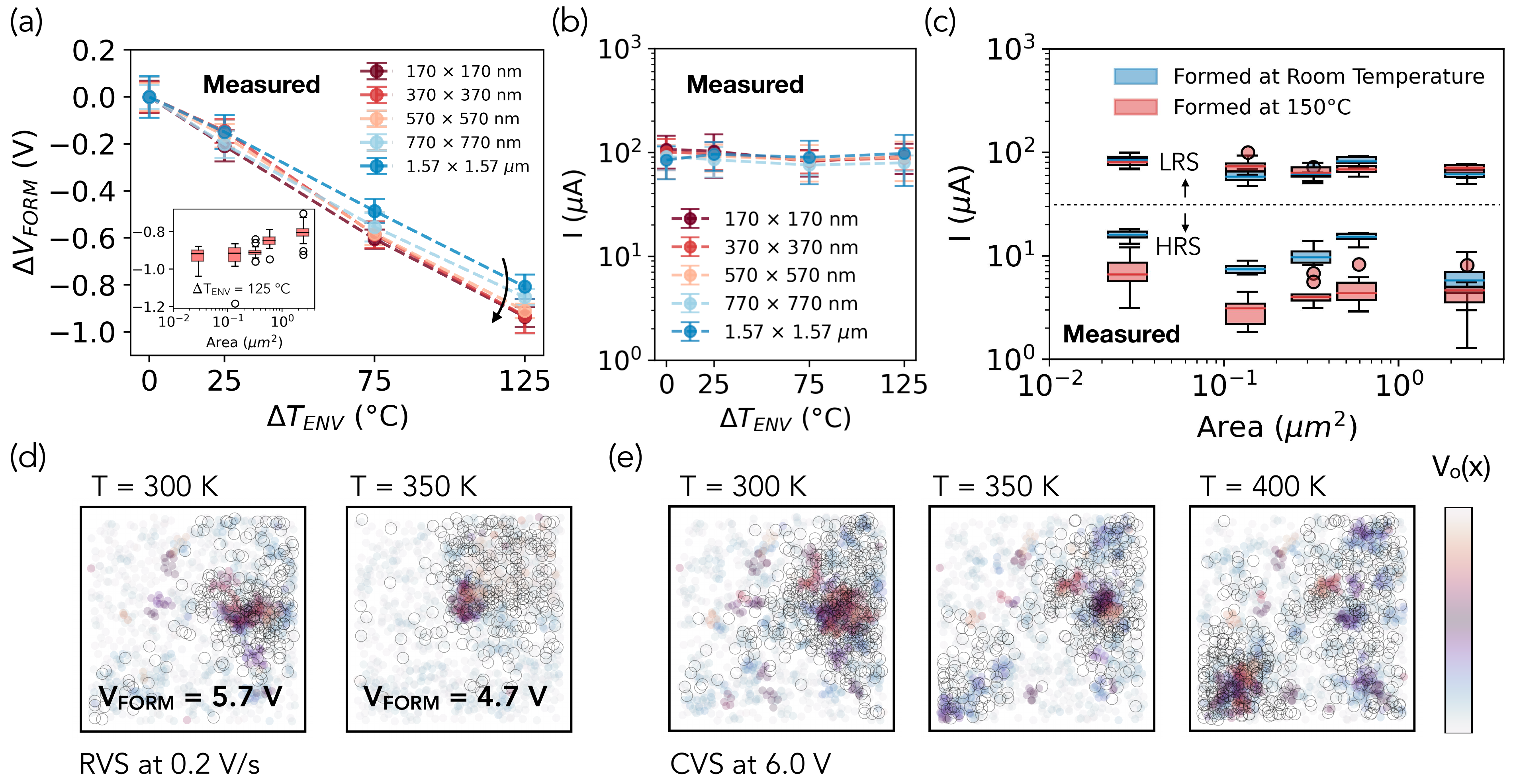}
        \caption{\textbf{Electroforming of the TiN/a-HfO$_x$/Ti/TiN device under increased ambient  temperature.} \textbf{(a)} Reduction in V$_{FORM}$ for the standard stack at different T$_{env}$ ranging from room temperature ($\Delta$T$_{env}$ = 0 °C) to 150 °C ($\Delta$T$_{env}$ = 125 °C). The data corresponding to $\Delta$T$_{env}$ = 125 °C is additionally plotted as a function of device area in the inset. \textbf{(b)} LRS currents measured at V$_{read}$ = 0.2 V for devices formed at increasing T$_{env}$. Error bars for each data point in (a)-(b) represent the standard deviation over 18 individual devices of a specified area. \textbf{(c)} High (LRS) and low (HRS) currents read at a voltage of V$_{SET}$/2, for devices initially electroformed either at room temperature or at 150 °C. Each box represents cycle-to-cycle variation over 10 consecutive switching cycles on the same measured device. \textbf{(d-e)} Simulated top-view vacancy profiles of the 7.5 $\times$ 7.5 nm\textsuperscript{2} device, under \textbf{(d)} RVS or \textbf{(e)} Constant Voltage Stress (CVS) electroforming at different temperatures. The V$_o$ are colored according to their vertical depth in the HfO$_x$ layer. The positions of migrating oxygen atoms are shown in black outlines.}
        \label{fig:temperature}
\end{figure}

In \textbf{Fig. \ref{fig:sweeprate}}, thermal effects were induced by internally-generated Joule heating at high voltages. The ambient temperature (T$_{env}$), however, remained constant. We now consider the effects of increasing T$_{env}$, which has been previously suggested as a method to reduce V$_{FORM}$ \cite{Su2018}. In \textbf{Fig. \ref{fig:temperature}(a)} we show the measured decrease in V$_{FORM}$ with increased temperature. The electroforming process is performed with the chip placed on top of a thermal wafer chuck. In this case, T$_{env}$ corresponds to the temperature of the thermal chuck, although the actual environment temperature at the device-level may be slightly lower. The data is shown with respect to reference V$_{FORM}$ at room temperature (T$_{env}$ = 25 °C). We see that V$_{FORM}$ is reduced almost linearly by $\sim$1 V for a 125 °C increase in T$_{env}$. The reduction of V$_{FORM}$ is most pronounced for the devices with smaller areas, as indicated by the black curved arrow and in the inset of \textbf{Fig. \ref{fig:temperature}(a)}, because they have less volume to dissipate the accumulated heat. Similarly to the case of varying sweep rates, we do not see any effect of the V$_{FORM}$ shift on I$_{LRS}$ under a fixed I$_{CC}$ (\textbf{Fig. \ref{fig:temperature}(b)}). However, the dynamic range of the resulting devices moderately increases at higher electroforming temperature (\textbf{Fig. \ref{fig:temperature}(c)}). While the LRS currents remain constant due to the the fixed I$_{CC}$, their HRS counterparts slightly decrease in all devices measured. This effect is solely attributed to the addition of heat during the actual forming process, as exposing already-formed devices to temperatures of 150 °C does not noticeably alter their switching properties (Supporting Information, \textbf{Fig. S4(c)}).

We simulate the effect of varying $T_{env}$ by uniformly increasing the initial temperature across the oxide, while fixing this parameter in the contacts to T$_{env}$. For the 7.5 $\times$ 7.5 nm\textsuperscript{2} device, a 50 °C increase in T$_{env}$ is sufficient to reduce V$_{FORM}$ by $\sim$1 V. The fact that a lower T$_{env}$ is needed to reach such a change in V$_{FORM}$ is a consequence of the smaller volume of the simulated devices, following the trend present in \textbf{Fig. \ref{fig:temperature}(a)}, and of the fact that T$_{env}$ is enforced in the contact regions. The structure of the conductive filament formed is only slightly narrower (\textbf{Fig. \ref{fig:temperature}(d)}). For comparison, we investigate the trend in T$_{env}$ under a Constant Voltage Stress (CVS) measurement procedure in \textbf{Fig. \ref{fig:temperature}(e)}. Here, V$_{FORM}$ is fixed to 6.0 V and the time to reach I$_{CC}$ varies. Increasing T$_{env}$ at V$_{FORM}$=6.0 V results in a combination of accelerated migration processes and increased Joule heating. The latter encourages lateral, thermally-driven diffusion of oxygen vacancies, which favors the formation of multiple, small filaments rather than a single one. These results are similar to the dispersed conductive filaments seen at higher sweep rates in \textbf{Fig. \ref{fig:sweeprate}(d)}.

Together, these results paint a consistent picture of the influence of T$_{env}$ and internal Joule heating on the formed filament morphologies. Increasing the temperature non-uniformly through Joule heating encourages lateral diffusion of V$_{o}$, leading to the appearance of multiple narrow conductive pathways and reduced dynamic range, as also shown in \textbf{Ref. \citenum{Su2018}}. This finding is consistent with the theorized wider filaments and more analog operation of VCM cells under optical stimulation \cite{Portner2021}, or implementing a thermal confinement layer \cite{Wu2017}. Both approaches enhance heating preferentially through the existing conductive pathways. While possibly useful to operate in the analog regime, these filament morphologies are not favorable if the goal is to obtain high resistance contrast for binary operation. Meanwhile, RVS forming procedures under high T$_{env}$ shift V$_{FORM}$ to lower values by uniformly accelerating the generation/migration processes involved. At these lower voltages, the effects of Joule heating on the forming process are diminished, leading to narrower, more directional conductive filaments. The imposed I$_{CC}$ then modulates the lateral spread of the conductive filament near the oxygen reservoir layer. As this region is oxidized during subsequent RESET operations, narrower filaments are left behind in the bulk of the oxide, which exhibit higher resistance in the HRS. Increasing T$_{env}$ through high-temperature forming protocols can thus be used to effectively harness the reduction in V$_{FORM}$ caused by adding heat, while avoiding more dispersed filament geometries.

\section{Conclusion}

Reduced stoichiometry and thermal engineering techniques have been known to influence the formation of conductive pathways in RRAM devices, but the ways in which they affect the kinetics of electroforming and the resulting filament morphology had remained unclear. In this work, we provide a deeper understanding on how these methods work by combining experimental characterizations of a TiN/a-HfO$_x$/Ti/TiN RRAM stack with atomistic modeling via KMC, thus bridging the gap between microscopic atomic rearrangements and device operation. By operating directly on an atomistic graph, our simulations can capture a wide range of the trends observed experimentally and native to filamentary RRAMs, such as the different current flow properties in the as-deposited and electroformed states, reduction in V$_{FORM}$ with larger device areas, and increase in V$_{FORM}$ with faster sweep rates and/or lower environment temperatures. In each case, we visualized the conductive filaments formed under different values of the investigated parameters. We observe a transition from vertical to lateral forming kinetics with reduced oxide stoichiometry, disperse, multi-filament growth encouraged by increased Joule heating, and uniformly accelerated growth favored by higher T$_{env}$. While both of the latter thermal effects have been observed in previous works under different conditions, the influence of uniform and non-uniform heat generation had not been previously distinguished. The different filament morphologies resulting from these situations can be connected to variations in the achievable HRS/LRS ratios; while reducing V$_{FORM}$ through stoichiometric-tuning or increased Joule heating can degrade the dynamic range during subsequent SET/RESET operations, electroforming at higher temperatures enhances it. The methodology developed here, although yielding insights specific to the HfO$_x$/Ti material stack, can be adapted to investigate valence-change-based resistive switching mechanisms in other binary oxide systems. Overall, our results provide a unified, atomistic understanding of the electroforming processes in RRAM and how they depend on available experimental parameters, as well as concrete insight into optimizing this crucial step to bring RRAM to technological readiness. 

\section{Methods}

\subsection{Device Fabrication and Characterization}

\subsubsection{Deposition of the device layer stack}

A 20 nm thick TiN bottom electrode (BE), followed by a 6 nm thick a-HfO$_x$ were first deposited by Plasma-Enhanced Atomic Layer Deposition (PEALD) at 300 $^{\circ}$C, using a tetrakis-(dimethylamino)titanium (TDMAT) precursor and (N$_2$, H$_2$) plasma. This was done without breaking the vacuum to prevent oxidation of the TiN. The Ti (10 nm)/ TiN (20 nm) /W (50 nm) were deposited by DC (Ti, W) and RF (TiN) sputtering in mixed (Ar, N$_2$) plasma and encapsulated by a 130 nm thick SiN passivation layer. After opening vias to access the devices, a 150 nm top electrode (TE) was sputtered. Further details can be found elsewhere \cite{Falcone2023}, and a schematic of the specific fabrication process can be found in the Supporting Information of \textbf{Ref. \citenum{portner2024actor}}. To further enhance the concentration of V$_o$ in the as-grown material, the O$_2$ plasma time was reduced from 10 to 1 s, which led to a reduction of the film density for the Leaky HfO$_x$ oxide, as confirmed by XRR measurements shown in the Supporting Information (\textbf{Fig. S1}). More details can be found in \textbf{Ref. \citenum{Stecconi2024}}.

\subsubsection{Stack Patterning}
After deposition of the TiN/a-HfO$_x$/Ti/TiN/W stack, the top three layers (Ti/TiN/W) are etched using Inductively Coupled Plasma Etching (ICP RIE) through a CHF\textsubscript{3}/SF\textsubscript{6}/N\textsubscript{2}-based etchant. The etch defines the active switching area. The active area is then encapsulated by 30 nm thick SiN, deposited via Plasma-Enhanced Chemical Vapour Deposition (PECVD) process at 300 $^{\circ}$C. The bottom two layers (a-HfO$_x$/TiN) are then etched to define the bottom electrode. First, the SiN encapsulation layer is etched through CHF\textsubscript{3}/O\textsubscript{2}-based reactive ion etching (RIE). The standard or leaky a-HfO$_x$ layer is then etched by ICP RIE through a CF\textsubscript{4}/Ar-based etchant and the TiN layer is patterned through SF\textsubscript{6}/N\textsubscript{2}-based RIE. Via openings to provide access to the memristor device are then done at both the TE and BE. For the former, SiN is etched, while both SiN and a-HfO$_x$ are etched for the latter. The vias are then filled by sputtering 150 nm of W, which is then patterning through SF\textsubscript{6}/N\textsubscript{2}-based RIE. A constant over-etch of 30 nm for the Ti/TiN layer, as measured though SEM imaging, was subtracted by the lateral dimensions (L$_{OX}$) during data post-processing.


\subsubsection{Structural Characterization by X-Ray Reflectivity}
To analyze the structural properties of the stacks, such as phase, thickness, density and interface quality between layers, XRR measurements were performed in a Bruker D8 discover diffractometer equipped with a rotating anode generator.
To obtain quantitative information on the various layer densities, thickness and the interfacial roughness, the XRR curves were analyzed by fitting a simulated curve based on a multilayer model to the measured data using the Leptos Reflectivity software \cite{Ulyanenkov2004}. The XRR fitting results and the corresponding extracted parameters are shown in the Supporting Information, \textbf{Fig. S1}.

\subsubsection{Device Measurement Setups}

The electrical measurements for \textbf{Figs. \ref{fig:stoichiometry}-\ref{fig:sweeprate}} were done using a B2912A Source Measure Unit (SMU) from Keysight. The temperature measurements in \textbf{Fig. \ref{fig:temperature}(a-b)} were carried out using a thermal wafer chuck and a B1500A Semiconductor Device Parameter Analyzer from Keysight. For all measurements, the bottom electrode (TiN) was grounded while the voltage was always applied to the top electrode (TiN/W). Electroforming was done with a current compliance of I$_{CC}$ = 20 \textmu A. In addition, a 10 $k\Omega$ external series resistor was used during the electroforming process to further ensure that any current overshoot during the forming is mitigated and prevent device breakdown. The series resistor was omitted for cycling and conductance readout measurements.

\subsection{Device Simulations}

\subsubsection{Simulated device structure generation}

Our d-KMC solver requires an input domain of `sites', each corresponding to the position of an atom or interstitial location in the device. To obtain atomic coordinates corresponding to amorphous HfO$_2$ structures with oxide dimensions of 2.5 $\times$ 2.5 nm\textsuperscript{2}, 5 $\times$ 5 nm\textsuperscript{2}, 7.5 $\times$ 7.5 nm\textsuperscript{2}, and 10 $\times$ 10 nm\textsuperscript{2}, we use the melt-quench procedure detailed in \textbf{Ref. \citenum{Kaniselvan2023}}. The identification of interstitial sites in such amorphous lattices is, however, non-trivial. To do this, we adopt a Voronoi approach \cite{Rycroft2009}. An initial Voronoi decomposition of the atomic structure gives us the locations of clustered vertices which corresponds to structural avoids in the amorphous lattice. We then collapse these clusters using a DBscan clustering method to a $\epsilon$ = 1.1 \AA, which can accurately reproduce the tetragonal and octahedral interstitial sites in the Ti/TiN contacts. The set of interstitial sites within the device defines the positions which can be occupied by diffusing oxygen ions.
 
\subsubsection{Activation Energy Calculations}

Activation energies for events, used as inputs for the KMC solver, are calculated using the NEB method\cite{neb} implemented in the CP2K code.\cite{cp2k} These energies are determined with the PBE functional,\cite{pbe} a localized DZVP basis set,\cite{dzvp} a charge density cutoff of 500 Ry, and a Gaussian-type orbital mapping cutoff of 60 Ry. The list of events and the corresponding activation energies are provided in Table S1. To identify saddle points, 7 images are used. Activation energies for events that occur in bulk HfO$_x$ (at the HfO$_x$/Ti interface) are averaged over 5 (3) different paths, while ion diffusion within Ti is evaluated for a single path between two octahedral sites.

\subsubsection{Driven Kinetic Monte Carlo solver}

We use a driven KMC application to model resistive switching on the device-scale while maintaining atomistic granularity of the underlying atomic migration processes. An initial version of the code was developed in \textbf{Ref. \citenum{Kaniselvan2023}}, computationally accelerated \cite{kaniselvan2024accelerated}, and extended in this work with treatment of Joule heating effects through an internal current flow model. The structural updates and solution of the site-resolved potential are unchanged from our implementation in the KMC model of \textbf{Ref. \citenum{Kaniselvan2023}}, and described there in detail. Here, we focus on the additional inclusion of the trap-assisted tunneling (TAT) current and Joule heating modules added subsequently to the application.

\noindent\textbf{Current flow} For every structural snapshot along the d-KMC timeline, we solve a semi-empirical model of current flow through the device. The current is defined between every two atomic sites in the device (excluding interstitial positions, and migrating oxygen ions). It is defined using a resistive network approach, where 

\begin{equation}
I_{i \rightarrow j} = -\frac{q^2}{\hbar} \cdot T_{ij} \cdot (\phi_i - \phi_j)
\label{eqn:current}
\end{equation}

Here, we can define a conductance matrix $\mathbf{G}$, where $G_{ij} = -q^2 / \hbar \cdot T_{ij}$. To parameterize the elements of $\mathbf{G}$, we consider that electronic transport inside the oxide mainly occurs by TAT through local defect states within the bandgap created by the resulting oxygen vacancies \cite{Kindsmller2018}. The bond-resolved transmission between atoms i and j, which carry this TAT effect, are modeled using Wentzel–Kramers–Brillouin (WKB) tunneling terms:

\begin{equation}
T = \exp\left(\frac{\sqrt{2m_e}}{\hbar}\cdot \int \sqrt{E(x)} dx\right)
\end{equation}

where $m_e$ = effective mass, and $E(x)$ is the potential barrier encountered during the tunneling process, stemming from a defect energy E$_d$. We consider the effect of varying potential barriers at different biases, which can correspond to direct- (low-field) or Fowler Nordheim (high-field) tunneling. Tunneling connections between contact sites and point defects are integrated over the range of energies defined by the applied potential $qV$. The value of T$_{ij}$ between atomic sites in the contacts is set to 1.

The effect of injection and extraction of current in this resistive network model is included through an effective source term, which appears as an extra two injection ($i$) and extraction ($e$) nodes in the resistive graph of $N$ sites. These are connected to the first/last layer of the contacts, respectively. The system of linear equations to solve then becomes:

\begin{equation}
   \mathbf{G}_{N\times N} \begin{bmatrix}
   \phi^{e} \\
   \phi^{i} \\
   \phi^{device}
   \end{bmatrix} =
   \begin{bmatrix}
   V_{app}\cdot G_{ie} \\
   -V_{app}\cdot G_{ei} \\
   0
   \end{bmatrix}
\end{equation}

Where $V_{app}$ is the voltage applied across the device. The last $N-2$ elements of the solution vector, $\phi_{device}$, contain `virtual potentials' which are used to calculate the pathways of bond-resolved current flow within the resistive-network approximation of the device, using \textbf{Eqn} \ref{eqn:current}. The final current through the device can be computed by summing all the connections from the injection node to the first layer of the contacts. The material parameters of the semi-empirical current solver (m$_e$ = 0.85m$_0$, E$_d$ = 1.6 eV) are calibrated by matching the computed current with that obtained by coherant Quantum Transport simulations across the pristine oxide across the range of relevant voltages (1-6 V) (Supporting Information, \textbf{Fig. S2(c)}).

\noindent\textbf{Model for Joule heating} Joule heating in RRAM occurs from heat dissipated in the oxide during electronic transitions. We capture this effect in simulations by computing the dissipated power $q_i$ at each site using:

\begin{equation} 
q_i = \alpha \sum_{j\neq i}I_{i \rightarrow j}\cdot (V_j - V_i)
\end{equation}

\noindent with a dissipation fraction $\alpha$ of 0.10 \cite{Ducry2020} which models the fraction of energy loss during electronic transitions which can be absorbed by the lattice as heat. This dissipated power is then used as a source term in the Fourier Heat equation:

\begin{equation}
    C_p \frac{dT}{dt} = k_{th}\nabla^2 T + q_i
\end{equation}

\noindent where the volumetric heat capacitance $C_p$ is set to 1.92 J/Kcm$^3$ \cite{Shen2023}, and the thermal conductivity k$_{th}$ is set to 11.4 W/mK for the contacts (Ti) and 1.0 W/mK for HfO$_2$ \cite{Panzer2009, Scott2018}. When solving this equation numerically on the atomic graph, we fix the temperature in the contacts to be 300 K as boundary conditions. We can thus compute a value of the effective `temperature' at every site, which is coupled into the rates computed for the next Kinetic Monte Carlo step.

\begin{acknowledgement}

This research is funded by the European Union and Swiss State Secretariat for Education, Research, and Innovation (SERI) within the PHASTRAC (grantID:101092096) project, by the Swiss State Secretariat for Education, Research, and Innovation (SERI) under the SwissChips Initiative, and by the Swiss National Science Foundation (SNSF) under ALMOND (grantID:198612). The authors acknowledge the Binnig and Rohrer Nanotechnology Center (BRNC) at IBM Research Europe - Zurich. MK acknowledges the NSERC Postgraduate Scholarship. Computational resources were provided by the Swiss National Supercomputing Center (CSCS) under projects s1119 and lp16. MK and KP contributed equally to this work.

\end{acknowledgement}

\begin{suppinfo}

The Supporting Information is available at [link], and includes details about the d-KMC model, XRR measurements, device characterization, area-(in)dependance of the SET voltage, and dynamic range measurements under different thermal conditions as mentioned in the main text.
\end{suppinfo}

\bibliography{manuscript}

\providecommand{\latin}[1]{#1}
\makeatletter
\providecommand{\doi}
  {\begingroup\let\do\@makeother\dospecials
  \catcode`\{=1 \catcode`\}=2 \doi@aux}
\providecommand{\doi@aux}[1]{\endgroup\texttt{#1}}
\makeatother
\providecommand*\mcitethebibliography{\thebibliography}
\csname @ifundefined\endcsname{endmcitethebibliography}  {\let\endmcitethebibliography\endthebibliography}{}
\begin{mcitethebibliography}{69}
\providecommand*\natexlab[1]{#1}
\providecommand*\mciteSetBstSublistMode[1]{}
\providecommand*\mciteSetBstMaxWidthForm[2]{}
\providecommand*\mciteBstWouldAddEndPuncttrue
  {\def\EndOfBibitem{\unskip.}}
\providecommand*\mciteBstWouldAddEndPunctfalse
  {\let\EndOfBibitem\relax}
\providecommand*\mciteSetBstMidEndSepPunct[3]{}
\providecommand*\mciteSetBstSublistLabelBeginEnd[3]{}
\providecommand*\EndOfBibitem{}
\mciteSetBstSublistMode{f}
\mciteSetBstMaxWidthForm{subitem}{(\alph{mcitesubitemcount})}
\mciteSetBstSublistLabelBeginEnd
  {\mcitemaxwidthsubitemform\space}
  {\relax}
  {\relax}

\bibitem[Pi \latin{et~al.}(2018)Pi, Li, Jiang, Xia, Xin, Yang, and Xia]{Pi2018}
Pi,~S.; Li,~C.; Jiang,~H.; Xia,~W.; Xin,~H.; Yang,~J.~J.; Xia,~Q. Memristor Crossbar Arrays with 6-nm Half-Pitch and 2-nm Critical Dimension. \emph{Nat. Nanotechnol.} \textbf{2018}, \emph{14}, 35--39\relax
\mciteBstWouldAddEndPuncttrue
\mciteSetBstMidEndSepPunct{\mcitedefaultmidpunct}
{\mcitedefaultendpunct}{\mcitedefaultseppunct}\relax
\EndOfBibitem
\bibitem[Yoon \latin{et~al.}(2019)Yoon, Kim, and Hwang]{Yoon2019}
Yoon,~K.~J.; Kim,~Y.; Hwang,~C.~S. What Will Come After V‐NAND—Vertical Resistive Switching Memory? \emph{Advanced Electronic Materials} \textbf{2019}, \emph{5}\relax
\mciteBstWouldAddEndPuncttrue
\mciteSetBstMidEndSepPunct{\mcitedefaultmidpunct}
{\mcitedefaultendpunct}{\mcitedefaultseppunct}\relax
\EndOfBibitem
\bibitem[Christensen \latin{et~al.}(2022)Christensen, Dittmann, Linares-Barranco, Sebastian, Le~Gallo, Redaelli, Slesazeck, Mikolajick, Spiga, Menzel, Valov, Milano, Ricciardi, Liang, Miao, Lanza, Quill, Keene, Salleo, Grollier, Marković, Mizrahi, Yao, Yang, Indiveri, Strachan, Datta, Vianello, Valentian, Feldmann, Li, Pernice, Bhaskaran, Furber, Neftci, Scherr, Maass, Ramaswamy, Tapson, Panda, Kim, Tanaka, Thorpe, Bartolozzi, Cleland, Posch, Liu, Panuccio, Mahmud, Mazumder, Hosseini, Mohsenin, Donati, Tolu, Galeazzi, Christensen, Holm, Ielmini, and Pryds]{Christensen2022}
Christensen,~D.~V.; Dittmann,~R.; Linares-Barranco,~B.; Sebastian,~A.; Le~Gallo,~M.; Redaelli,~A.; Slesazeck,~S.; Mikolajick,~T.; Spiga,~S.; Menzel,~S.; Valov,~I.; Milano,~G.; Ricciardi,~C.; Liang,~S.-J.; Miao,~F.; Lanza,~M.; Quill,~T.~J.; Keene,~S.~T.; Salleo,~A.; Grollier,~J.; Marković,~D.; Mizrahi,~A.; Yao,~P.; Yang,~J.~J.; Indiveri,~G.; Strachan,~J.~P.; Datta,~S.; Vianello,~E.; Valentian,~A.; Feldmann,~J.; Li,~X.; Pernice,~W. H.~P.; Bhaskaran,~H.; Furber,~S.; Neftci,~E.; Scherr,~F.; Maass,~W.; Ramaswamy,~S.; Tapson,~J.; Panda,~P.; Kim,~Y.; Tanaka,~G.; Thorpe,~S.; Bartolozzi,~C.; Cleland,~T.~A.; Posch,~C.; Liu,~S.; Panuccio,~G.; Mahmud,~M.; Mazumder,~A.~N.; Hosseini,~M.; Mohsenin,~T.; Donati,~E.; Tolu,~S.; Galeazzi,~R.; Christensen,~M.~E.; Holm,~S.; Ielmini,~D.; Pryds,~N. 2022 roadmap on neuromorphic computing and engineering. \emph{Neuromorphic Computing and Engineering} \textbf{2022}, \emph{2}, 022501\relax
\mciteBstWouldAddEndPuncttrue
\mciteSetBstMidEndSepPunct{\mcitedefaultmidpunct}
{\mcitedefaultendpunct}{\mcitedefaultseppunct}\relax
\EndOfBibitem
\bibitem[Ielmini and Wong(2018)Ielmini, and Wong]{Ielmini2018}
Ielmini,~D.; Wong,~H.-S.~P. In-Memory Computing with Resistive Switching Devices. \emph{Nat. Electron.} \textbf{2018}, \emph{1}, 333--343\relax
\mciteBstWouldAddEndPuncttrue
\mciteSetBstMidEndSepPunct{\mcitedefaultmidpunct}
{\mcitedefaultendpunct}{\mcitedefaultseppunct}\relax
\EndOfBibitem
\bibitem[Dittmann \latin{et~al.}(2021)Dittmann, Menzel, and Waser]{Dittmann2021}
Dittmann,~R.; Menzel,~S.; Waser,~R. Nanoionic memristive phenomena in metal oxides: the valence change mechanism. \emph{Advances in Physics} \textbf{2021}, \emph{70}, 155–349\relax
\mciteBstWouldAddEndPuncttrue
\mciteSetBstMidEndSepPunct{\mcitedefaultmidpunct}
{\mcitedefaultendpunct}{\mcitedefaultseppunct}\relax
\EndOfBibitem
\bibitem[Chen(2013)]{Chen2013}
Chen,~A. Forming voltage scaling of resistive switching memories. 71st Device Research Conference. 2013\relax
\mciteBstWouldAddEndPuncttrue
\mciteSetBstMidEndSepPunct{\mcitedefaultmidpunct}
{\mcitedefaultendpunct}{\mcitedefaultseppunct}\relax
\EndOfBibitem
\bibitem[An \latin{et~al.}(2024)An, Yan, Yeom, Jeong, Eadi, Lee, and Kwon]{An2024}
An,~Y.-J.; Yan,~H.; Yeom,~C.-m.; Jeong,~J.-k.; Eadi,~S.~B.; Lee,~H.-D.; Kwon,~H.-M. Effects of thermal annealing on analog resistive switching behavior in bilayer HfO$_2$/ZnO synaptic devices: the role of ZnO grain boundaries. \emph{Nanoscale} \textbf{2024}, \emph{16}, 4609–4619\relax
\mciteBstWouldAddEndPuncttrue
\mciteSetBstMidEndSepPunct{\mcitedefaultmidpunct}
{\mcitedefaultendpunct}{\mcitedefaultseppunct}\relax
\EndOfBibitem
\bibitem[Xue \latin{et~al.}(2012)Xue, Chen, Wang, Zhou, Chang, Fowler, and Lee]{Xue2012}
Xue,~F.; Chen,~Y.-T.; Wang,~Y.; Zhou,~F.; Chang,~Y.-F.; Fowler,~B.; Lee,~J. The Effect of Plasma Treatment on Reducing Electroforming Voltage of Silicon Oxide RRAM. \emph{ECS Transactions} \textbf{2012}, \emph{45}, 245–250\relax
\mciteBstWouldAddEndPuncttrue
\mciteSetBstMidEndSepPunct{\mcitedefaultmidpunct}
{\mcitedefaultendpunct}{\mcitedefaultseppunct}\relax
\EndOfBibitem
\bibitem[Liu \latin{et~al.}(2010)Liu, Long, Lv, Wang, Niu, Huo, Chen, and Liu]{Liu2010}
Liu,~Q.; Long,~S.; Lv,~H.; Wang,~W.; Niu,~J.; Huo,~Z.; Chen,~J.; Liu,~M. Controllable Growth of Nanoscale Conductive Filaments in Solid-Electrolyte-Based ReRAM by Using a Metal Nanocrystal Covered Bottom Electrode. \emph{ACS Nano} \textbf{2010}, \emph{4}, 6162–6168\relax
\mciteBstWouldAddEndPuncttrue
\mciteSetBstMidEndSepPunct{\mcitedefaultmidpunct}
{\mcitedefaultendpunct}{\mcitedefaultseppunct}\relax
\EndOfBibitem
\bibitem[Niu \latin{et~al.}(2016)Niu, Calka, Auf~der Maur, Santoni, Guha, Fraschke, Hamoumou, Gautier, Perez, Walczyk, Wenger, Di~Carlo, Alff, and Schroeder]{Niu2016}
Niu,~G.; Calka,~P.; Auf~der Maur,~M.; Santoni,~F.; Guha,~S.; Fraschke,~M.; Hamoumou,~P.; Gautier,~B.; Perez,~E.; Walczyk,~C.; Wenger,~C.; Di~Carlo,~A.; Alff,~L.; Schroeder,~T. Geometric conductive filament confinement by nanotips for resistive switching of HfO$_2$-RRAM devices with high performance. \emph{Scientific Reports} \textbf{2016}, \emph{6}\relax
\mciteBstWouldAddEndPuncttrue
\mciteSetBstMidEndSepPunct{\mcitedefaultmidpunct}
{\mcitedefaultendpunct}{\mcitedefaultseppunct}\relax
\EndOfBibitem
\bibitem[Jeon \latin{et~al.}(2023)Jeon, Eom, Lee, Kim, and Lee]{Jeon2023}
Jeon,~J.; Eom,~K.; Lee,~M.; Kim,~S.; Lee,~H. Collective Control of Potential-Constrained Oxygen Vacancies in Oxide Heterostructures for Gradual Resistive Switching. \emph{Small} \textbf{2023}, 2301452\relax
\mciteBstWouldAddEndPuncttrue
\mciteSetBstMidEndSepPunct{\mcitedefaultmidpunct}
{\mcitedefaultendpunct}{\mcitedefaultseppunct}\relax
\EndOfBibitem
\bibitem[Sarantopoulos \latin{et~al.}(2024)Sarantopoulos, Lange, Rivadulla, Menzel, and Dittmann]{Sarantopoulos2024}
Sarantopoulos,~A.; Lange,~K.; Rivadulla,~F.; Menzel,~S.; Dittmann,~R. Resistive Switching Acceleration Induced by Thermal Confinement. \emph{Advanced Electronic Materials} \textbf{2024}, \relax
\mciteBstWouldAddEndPunctfalse
\mciteSetBstMidEndSepPunct{\mcitedefaultmidpunct}
{}{\mcitedefaultseppunct}\relax
\EndOfBibitem
\bibitem[Wu \latin{et~al.}(2017)Wu, Wu, Gao, Deng, Yu, and Qian]{Wu2017}
Wu,~W.; Wu,~H.; Gao,~B.; Deng,~N.; Yu,~S.; Qian,~H. Improving Analog Switching in HfO$_2$-Based Resistive Memory With a Thermal Enhanced Layer. \emph{IEEE Electron Device Letters} \textbf{2017}, \emph{38}, 1019–1022\relax
\mciteBstWouldAddEndPuncttrue
\mciteSetBstMidEndSepPunct{\mcitedefaultmidpunct}
{\mcitedefaultendpunct}{\mcitedefaultseppunct}\relax
\EndOfBibitem
\bibitem[Jeong \latin{et~al.}(2015)Jeong, Kim, and Lu]{Jeong2015}
Jeong,~Y.; Kim,~S.; Lu,~W.~D. Utilizing multiple state variables to improve the dynamic range of analog switching in a memristor. \emph{Applied Physics Letters} \textbf{2015}, \emph{107}\relax
\mciteBstWouldAddEndPuncttrue
\mciteSetBstMidEndSepPunct{\mcitedefaultmidpunct}
{\mcitedefaultendpunct}{\mcitedefaultseppunct}\relax
\EndOfBibitem
\bibitem[Portner \latin{et~al.}(2021)Portner, Schmuck, Lehmann, Weilenmann, Haffner, Ma, Leuthold, Luisier, and Emboras]{Portner2021}
Portner,~K.; Schmuck,~M.; Lehmann,~P.; Weilenmann,~C.; Haffner,~C.; Ma,~P.; Leuthold,~J.; Luisier,~M.; Emboras,~A. Analog Nanoscale Electro-Optical Synapses for Neuromorphic Computing Applications. \emph{ACS Nano} \textbf{2021}, \emph{15}, 14776–14785\relax
\mciteBstWouldAddEndPuncttrue
\mciteSetBstMidEndSepPunct{\mcitedefaultmidpunct}
{\mcitedefaultendpunct}{\mcitedefaultseppunct}\relax
\EndOfBibitem
\bibitem[Sun \latin{et~al.}(2019)Sun, Gao, Chi, Xia, Yang, Qian, and Wu]{Sun2019}
Sun,~W.; Gao,~B.; Chi,~M.; Xia,~Q.; Yang,~J.~J.; Qian,~H.; Wu,~H. Understanding memristive switching via in situ characterization and device modeling. \emph{Nature Communications} \textbf{2019}, \emph{10}\relax
\mciteBstWouldAddEndPuncttrue
\mciteSetBstMidEndSepPunct{\mcitedefaultmidpunct}
{\mcitedefaultendpunct}{\mcitedefaultseppunct}\relax
\EndOfBibitem
\bibitem[Urquiza \latin{et~al.}(2021)Urquiza, Islam, van Duin, Cartoix{\`{a}}, and Strachan]{Urquiza2021}
Urquiza,~M.~L.; Islam,~M.~M.; van Duin,~A. C.~T.; Cartoix{\`{a}},~X.; Strachan,~A. Atomistic Insights on the Full Operation Cycle of a HfO$_2$-Based Resistive Random Access Memory Cell from Molecular Dynamics. \emph{{ACS} Nano} \textbf{2021}, \emph{15}, 12945--12954\relax
\mciteBstWouldAddEndPuncttrue
\mciteSetBstMidEndSepPunct{\mcitedefaultmidpunct}
{\mcitedefaultendpunct}{\mcitedefaultseppunct}\relax
\EndOfBibitem
\bibitem[Kaniselvan \latin{et~al.}(2023)Kaniselvan, Luisier, and Mladenović]{Kaniselvan2023}
Kaniselvan,~M.; Luisier,~M.; Mladenović,~M. An Atomistic Model of Field-Induced Resistive Switching in Valence Change Memory. \emph{ACS Nano} \textbf{2023}, \emph{17}, 8281–8292\relax
\mciteBstWouldAddEndPuncttrue
\mciteSetBstMidEndSepPunct{\mcitedefaultmidpunct}
{\mcitedefaultendpunct}{\mcitedefaultseppunct}\relax
\EndOfBibitem
\bibitem[Woo \latin{et~al.}(2024)Woo, Williams, and Kumar]{Woo2024}
Woo,~K.~S.; Williams,~R.~S.; Kumar,~S. Localized Conduction Channels in Memristors. \emph{Chemical Reviews} \textbf{2024}, \emph{125}, 294–325\relax
\mciteBstWouldAddEndPuncttrue
\mciteSetBstMidEndSepPunct{\mcitedefaultmidpunct}
{\mcitedefaultendpunct}{\mcitedefaultseppunct}\relax
\EndOfBibitem
\bibitem[Zhang \latin{et~al.}(2022)Zhang, Ren, Ganesh, and Cao]{Zhang2022}
Zhang,~K.; Ren,~Y.; Ganesh,~P.; Cao,~Y. Effect of electrode and oxide properties on the filament kinetics during electroforming in metal-oxide-based memories. \emph{npj Computational Materials} \textbf{2022}, \emph{8}\relax
\mciteBstWouldAddEndPuncttrue
\mciteSetBstMidEndSepPunct{\mcitedefaultmidpunct}
{\mcitedefaultendpunct}{\mcitedefaultseppunct}\relax
\EndOfBibitem
\bibitem[Aeschlimann \latin{et~al.}(2023)Aeschlimann, Bani-Hashemian, Ducry, Emboras, and Luisier]{Aeschlimann2023}
Aeschlimann,~J.; Bani-Hashemian,~M.; Ducry,~F.; Emboras,~A.; Luisier,~M. Insights into few-atom conductive bridging random access memory cells with a combined force-field/ab initio scheme. \emph{Solid-State Electronics} \textbf{2023}, \emph{199}, 108493\relax
\mciteBstWouldAddEndPuncttrue
\mciteSetBstMidEndSepPunct{\mcitedefaultmidpunct}
{\mcitedefaultendpunct}{\mcitedefaultseppunct}\relax
\EndOfBibitem
\bibitem[Celano \latin{et~al.}(2014)Celano, Goux, Belmonte, Opsomer, Franquet, Schulze, Detavernier, Richard, Bender, Jurczak, and Vandervorst]{Celano2014}
Celano,~U.; Goux,~L.; Belmonte,~A.; Opsomer,~K.; Franquet,~A.; Schulze,~A.; Detavernier,~C.; Richard,~O.; Bender,~H.; Jurczak,~M.; Vandervorst,~W. Three-Dimensional Observation of the Conductive Filament in Nanoscaled Resistive Memory Devices. \emph{Nano Letters} \textbf{2014}, \emph{14}, 2401–2406\relax
\mciteBstWouldAddEndPuncttrue
\mciteSetBstMidEndSepPunct{\mcitedefaultmidpunct}
{\mcitedefaultendpunct}{\mcitedefaultseppunct}\relax
\EndOfBibitem
\bibitem[Celano \latin{et~al.}(2015)Celano, Goux, Degraeve, Fantini, Richard, Bender, Jurczak, and Vandervorst]{Celano2015}
Celano,~U.; Goux,~L.; Degraeve,~R.; Fantini,~A.; Richard,~O.; Bender,~H.; Jurczak,~M.; Vandervorst,~W. Imaging the Three-Dimensional Conductive Channel in Filamentary-Based Oxide Resistive Switching Memory. \emph{Nano Letters} \textbf{2015}, \emph{15}, 7970–7975\relax
\mciteBstWouldAddEndPuncttrue
\mciteSetBstMidEndSepPunct{\mcitedefaultmidpunct}
{\mcitedefaultendpunct}{\mcitedefaultseppunct}\relax
\EndOfBibitem
\bibitem[Ma \latin{et~al.}(2020)Ma, Yeoh, Shen, Goodwill, Bain, and Skowronski]{Ma2020}
Ma,~Y.; Yeoh,~P.~P.; Shen,~L.; Goodwill,~J.~M.; Bain,~J.~A.; Skowronski,~M. Evolution of the conductive filament with cycling in TaOx-based resistive switching devices. \emph{Journal of Applied Physics} \textbf{2020}, \emph{128}\relax
\mciteBstWouldAddEndPuncttrue
\mciteSetBstMidEndSepPunct{\mcitedefaultmidpunct}
{\mcitedefaultendpunct}{\mcitedefaultseppunct}\relax
\EndOfBibitem
\bibitem[Falcone \latin{et~al.}(2023)Falcone, Menzel, Stecconi, Porta, Carraria-Martinotti, Offrein, and Bragaglia]{Falcone2023}
Falcone,~D.~F.; Menzel,~S.; Stecconi,~T.; Porta,~A.~L.; Carraria-Martinotti,~L.; Offrein,~B.~J.; Bragaglia,~V. Physical modeling and design rules of analog Conductive Metal Oxide-HfO$_2$ ReRAM. 2023 IEEE International Memory Workshop (IMW). 2023; p 1–4\relax
\mciteBstWouldAddEndPuncttrue
\mciteSetBstMidEndSepPunct{\mcitedefaultmidpunct}
{\mcitedefaultendpunct}{\mcitedefaultseppunct}\relax
\EndOfBibitem
\bibitem[Stecconi \latin{et~al.}(2024)Stecconi, Bragaglia, Rasch, Carta, Horst, Falcone, ten Kate, Gong, Ando, Olziersky, and Offrein]{Stecconi2024}
Stecconi,~T.; Bragaglia,~V.; Rasch,~M.~J.; Carta,~F.; Horst,~F.; Falcone,~D.~F.; ten Kate,~S.~C.; Gong,~N.; Ando,~T.; Olziersky,~A.; Offrein,~B. Analog Resistive Switching Devices for Training Deep Neural Networks with the Novel Tiki-Taka Algorithm. \emph{Nano Letters} \textbf{2024}, \emph{24}, 866–872\relax
\mciteBstWouldAddEndPuncttrue
\mciteSetBstMidEndSepPunct{\mcitedefaultmidpunct}
{\mcitedefaultendpunct}{\mcitedefaultseppunct}\relax
\EndOfBibitem
\bibitem[Stecconi \latin{et~al.}(2022)Stecconi, Guido, Berchialla, La~Porta, Weiss, Popoff, Halter, Sousa, Horst, Dávila, Drechsler, Dittmann, Offrein, and Bragaglia]{Stecconi2022}
Stecconi,~T.; Guido,~R.; Berchialla,~L.; La~Porta,~A.; Weiss,~J.; Popoff,~Y.; Halter,~M.; Sousa,~M.; Horst,~F.; Dávila,~D.; Drechsler,~U.; Dittmann,~R.; Offrein,~B.~J.; Bragaglia,~V. Filamentary TaO$_x$/HfO$_2$ ReRAM Devices for Neural Networks Training with Analog In‐Memory Computing. \emph{Advanced Electronic Materials} \textbf{2022}, \emph{8}\relax
\mciteBstWouldAddEndPuncttrue
\mciteSetBstMidEndSepPunct{\mcitedefaultmidpunct}
{\mcitedefaultendpunct}{\mcitedefaultseppunct}\relax
\EndOfBibitem
\bibitem[Kaniselvan \latin{et~al.}(2024)Kaniselvan, Mladenović, Clarysse, Portner, and Luisier]{Kaniselvan2024}
Kaniselvan,~M.; Mladenović,~M.; Clarysse,~J.; Portner,~K.; Luisier,~M. Insights behind multi-level conductance transitions in HfO$_x$ memristors. 2024 Device Research Conference (DRC). 2024; p 1–2\relax
\mciteBstWouldAddEndPuncttrue
\mciteSetBstMidEndSepPunct{\mcitedefaultmidpunct}
{\mcitedefaultendpunct}{\mcitedefaultseppunct}\relax
\EndOfBibitem
\bibitem[Kaniselvan \latin{et~al.}(2024)Kaniselvan, Maeder, Mladenovi{\'c}, Luisier, and Ziogas]{kaniselvan2024accelerated}
Kaniselvan,~M.; Maeder,~A.; Mladenovi{\'c},~M.; Luisier,~M.; Ziogas,~A.~N. Accelerated Atomistic Kinetic Monte Carlo Simulations of Resistive Memory Arrays. 2024 SC24: International Conference for High Performance Computing, Networking, Storage and Analysis SC. 2024; pp 1450--1465\relax
\mciteBstWouldAddEndPuncttrue
\mciteSetBstMidEndSepPunct{\mcitedefaultmidpunct}
{\mcitedefaultendpunct}{\mcitedefaultseppunct}\relax
\EndOfBibitem
\bibitem[Govoreanu \latin{et~al.}(2011, pp. 31.6.1-31.6.4)Govoreanu, Kar, Chen, Paraschiv, Kubicek, Fantini, Radu, Goux, Clima, Degraeve, Jossart, Richard, Vandeweyer, Seo, Hendrickx, Pourtois, Bender, Altimime, Wouters, Kittl, and Jurczak]{Govoreanu2011}
Govoreanu,~B.; Kar,~G.; Chen,~Y.-Y.; Paraschiv,~V.; Kubicek,~S.; Fantini,~A.; Radu,~I.; Goux,~L.; Clima,~S.; Degraeve,~R.; Jossart,~N.; Richard,~O.; Vandeweyer,~T.; Seo,~K.; Hendrickx,~P.; Pourtois,~G.; Bender,~H.; Altimime,~L.; Wouters,~D.; Kittl,~J.; Jurczak,~M. 10x10nm Hf/HfO$_2$ crossbar resistive RAM with excellent performance, reliability and low-energy operation. IEEE IEDM. 2011, pp. 31.6.1-31.6.4\relax
\mciteBstWouldAddEndPuncttrue
\mciteSetBstMidEndSepPunct{\mcitedefaultmidpunct}
{\mcitedefaultendpunct}{\mcitedefaultseppunct}\relax
\EndOfBibitem
\bibitem[Li \latin{et~al.}(2014)Li, Ho, Lee, Chen, Hsu, Lu, Lin, Chen, Wu, Hou, Lin, Chen, Lai, Li, Yang, Wu, and Yang]{KaiShinLi2014}
Li,~K.-S.; Ho,~C.; Lee,~M.-T.; Chen,~M.-C.; Hsu,~C.-L.; Lu,~J.~M.; Lin,~C.~H.; Chen,~C.~C.; Wu,~B.~W.; Hou,~Y.~F.; Lin,~C.~Y.; Chen,~Y.~J.; Lai,~T.~Y.; Li,~M.~Y.; Yang,~I.; Wu,~C.~S.; Yang,~F.-L. Utilizing Sub-5 nm sidewall electrode technology for atomic-scale resistive memory fabrication. 2014 Symposium on VLSI Technology (VLSI-Technology): Digest of Technical Papers. 2014; p 1–2\relax
\mciteBstWouldAddEndPuncttrue
\mciteSetBstMidEndSepPunct{\mcitedefaultmidpunct}
{\mcitedefaultendpunct}{\mcitedefaultseppunct}\relax
\EndOfBibitem
\bibitem[Jónsson \latin{et~al.}(1998)Jónsson, Mills, and Jacobsen]{neb}
Jónsson,~H.; Mills,~G.; Jacobsen,~K.~W. Nudged elastic band method for finding minimum energy paths of transitions. Classical and Quantum Dynamics in Condensed Phase Simulations. 1998\relax
\mciteBstWouldAddEndPuncttrue
\mciteSetBstMidEndSepPunct{\mcitedefaultmidpunct}
{\mcitedefaultendpunct}{\mcitedefaultseppunct}\relax
\EndOfBibitem
\bibitem[O{\textquotesingle}Hara \latin{et~al.}(2014)O{\textquotesingle}Hara, Bersuker, and Demkov]{OHara2014}
O{\textquotesingle}Hara,~A.; Bersuker,~G.; Demkov,~A.~A. Assessing Hafnium on Hafnia as an Oxygen Getter. \emph{J. Appl. Phys.} \textbf{2014}, \emph{115}, 183703\relax
\mciteBstWouldAddEndPuncttrue
\mciteSetBstMidEndSepPunct{\mcitedefaultmidpunct}
{\mcitedefaultendpunct}{\mcitedefaultseppunct}\relax
\EndOfBibitem
\bibitem[Bertaud \latin{et~al.}(2012)Bertaud, Sowinska, Walczyk, Thiess, Gloskovskii, Walczyk, and Schroeder]{Bertaud2012}
Bertaud,~T.; Sowinska,~M.; Walczyk,~D.; Thiess,~S.; Gloskovskii,~A.; Walczyk,~C.; Schroeder,~T. In-operando and non-destructive analysis of the resistive switching in the Ti/HfO$_2$/TiN-based system by hard x-ray photoelectron spectroscopy. \emph{Applied Physics Letters} \textbf{2012}, \emph{101}, 143501\relax
\mciteBstWouldAddEndPuncttrue
\mciteSetBstMidEndSepPunct{\mcitedefaultmidpunct}
{\mcitedefaultendpunct}{\mcitedefaultseppunct}\relax
\EndOfBibitem
\bibitem[Privitera \latin{et~al.}(2013)Privitera, Bersuker, Butcher, Kalantarian, Lombardo, Bongiorno, Geer, Gilmer, and Kirsch]{Privitera2013-vc}
Privitera,~S.; Bersuker,~G.; Butcher,~B.; Kalantarian,~A.; Lombardo,~S.; Bongiorno,~C.; Geer,~R.; Gilmer,~D.~C.; Kirsch,~P.~D. Microscopy Study of the Conductive Filament in HfO$_2$ Resistive Switching Memory Devices. \emph{Microelectron. Eng.} \textbf{2013}, \emph{109}, 75--78\relax
\mciteBstWouldAddEndPuncttrue
\mciteSetBstMidEndSepPunct{\mcitedefaultmidpunct}
{\mcitedefaultendpunct}{\mcitedefaultseppunct}\relax
\EndOfBibitem
\bibitem[Hubbard \latin{et~al.}(2021)Hubbard, Lodico, Chan, Mecklenburg, and Regan]{Hubbard2021}
Hubbard,~W.~A.; Lodico,~J.~J.; Chan,~H.~L.; Mecklenburg,~M.; Regan,~B.~C. Imaging Dielectric Breakdown in Valence Change Memory. \emph{Advanced Functional Materials} \textbf{2021}, \emph{32}\relax
\mciteBstWouldAddEndPuncttrue
\mciteSetBstMidEndSepPunct{\mcitedefaultmidpunct}
{\mcitedefaultendpunct}{\mcitedefaultseppunct}\relax
\EndOfBibitem
\bibitem[Fang \latin{et~al.}(2014)Fang, Wang, Sohn, Weng, Zhang, Chen, Tang, Lo, Provine, Wong, Wong, and Kwong]{Fang2014}
Fang,~Z.; Wang,~X.~P.; Sohn,~J.; Weng,~B.~B.; Zhang,~Z.~P.; Chen,~Z.~X.; Tang,~Y.~Z.; Lo,~G.-Q.; Provine,~J.; Wong,~S.~S.; Wong,~H.-S.~P.; Kwong,~D.-L. The Role of Ti Capping Layer in HfO$_x$ Based RRAM Devices. \emph{IEEE Electron Device Letters} \textbf{2014}, \emph{35}, 912–914\relax
\mciteBstWouldAddEndPuncttrue
\mciteSetBstMidEndSepPunct{\mcitedefaultmidpunct}
{\mcitedefaultendpunct}{\mcitedefaultseppunct}\relax
\EndOfBibitem
\bibitem[Kamiya \latin{et~al.}(2013)Kamiya, Yang, Nagata, Park, Magyari-K\"{o}pe, Chikyow, Yamada, Niwa, Nishi, and Shiraishi]{Kamiya2013}
Kamiya,~K.; Yang,~M.~Y.; Nagata,~T.; Park,~S.-G.; Magyari-K\"{o}pe,~B.; Chikyow,~T.; Yamada,~K.; Niwa,~M.; Nishi,~Y.; Shiraishi,~K. Generalized mechanism of the resistance switching in binary-oxide-based resistive random-access memories. \emph{Physical Review B} \textbf{2013}, \emph{87}\relax
\mciteBstWouldAddEndPuncttrue
\mciteSetBstMidEndSepPunct{\mcitedefaultmidpunct}
{\mcitedefaultendpunct}{\mcitedefaultseppunct}\relax
\EndOfBibitem
\bibitem[Schmitt \latin{et~al.}(2017)Schmitt, Spring, Korobko, and Rupp]{Schmitt2017}
Schmitt,~R.; Spring,~J.; Korobko,~R.; Rupp,~J.~L. Design of Oxygen Vacancy Configuration for Memristive Systems. \emph{ACS Nano} \textbf{2017}, \emph{11}, 8881–8891\relax
\mciteBstWouldAddEndPuncttrue
\mciteSetBstMidEndSepPunct{\mcitedefaultmidpunct}
{\mcitedefaultendpunct}{\mcitedefaultseppunct}\relax
\EndOfBibitem
\bibitem[Clima \latin{et~al.}(2012)Clima, Chen, Degraeve, Mees, Sankaran, Govoreanu, Jurczak, De~Gendt, and Pourtois]{Clima2012}
Clima,~S.; Chen,~Y.~Y.; Degraeve,~R.; Mees,~M.; Sankaran,~K.; Govoreanu,~B.; Jurczak,~M.; De~Gendt,~S.; Pourtois,~G. First-principles simulation of oxygen diffusion in HfO$_x$: Role in the resistive switching mechanism. \emph{Applied Physics Letters} \textbf{2012}, \emph{100}\relax
\mciteBstWouldAddEndPuncttrue
\mciteSetBstMidEndSepPunct{\mcitedefaultmidpunct}
{\mcitedefaultendpunct}{\mcitedefaultseppunct}\relax
\EndOfBibitem
\bibitem[Xiao \latin{et~al.}(2019)Xiao, Yu, and Watanabe]{Xiao2019}
Xiao,~B.; Yu,~X.; Watanabe,~S. A Comparative Study on the Diffusion Behaviors of Metal and Oxygen Ions in Metal-Oxide-Based Resistance Switches via ab Initio Molecular Dynamics Simulations. \emph{ACS Applied Electronic Materials} \textbf{2019}, \emph{1}, 585–594\relax
\mciteBstWouldAddEndPuncttrue
\mciteSetBstMidEndSepPunct{\mcitedefaultmidpunct}
{\mcitedefaultendpunct}{\mcitedefaultseppunct}\relax
\EndOfBibitem
\bibitem[Tsurumaki-Fukuchi \latin{et~al.}(2023)Tsurumaki-Fukuchi, Katase, Ohta, Arita, and Takahashi]{TsurumakiFukuchi2023}
Tsurumaki-Fukuchi,~A.; Katase,~T.; Ohta,~H.; Arita,~M.; Takahashi,~Y. Direct Imaging of Ion Migration in Amorphous Oxide Electronic Synapses with Intrinsic Analog Switching Characteristics. \emph{ACS Applied Materials \& Interfaces} \textbf{2023}, \emph{15}, 16842–16852\relax
\mciteBstWouldAddEndPuncttrue
\mciteSetBstMidEndSepPunct{\mcitedefaultmidpunct}
{\mcitedefaultendpunct}{\mcitedefaultseppunct}\relax
\EndOfBibitem
\bibitem[Nauenheim \latin{et~al.}(2010)Nauenheim, Kuegeler, Ruediger, and Waser]{Nauenheim2010}
Nauenheim,~C.; Kuegeler,~C.; Ruediger,~A.; Waser,~R. Investigation of the electroforming process in resistively switching TiO$_2$ nanocrosspoint junctions. \emph{Applied Physics Letters} \textbf{2010}, \emph{96}\relax
\mciteBstWouldAddEndPuncttrue
\mciteSetBstMidEndSepPunct{\mcitedefaultmidpunct}
{\mcitedefaultendpunct}{\mcitedefaultseppunct}\relax
\EndOfBibitem
\bibitem[Stathopoulos \latin{et~al.}(2019)Stathopoulos, Michalas, Khiat, Serb, and Prodromakis]{Stathopoulos2019}
Stathopoulos,~S.; Michalas,~L.; Khiat,~A.; Serb,~A.; Prodromakis,~T. An Electrical Characterisation Methodology for Benchmarking Memristive Device Technologies. \emph{Scientific Reports} \textbf{2019}, \emph{9}\relax
\mciteBstWouldAddEndPuncttrue
\mciteSetBstMidEndSepPunct{\mcitedefaultmidpunct}
{\mcitedefaultendpunct}{\mcitedefaultseppunct}\relax
\EndOfBibitem
\bibitem[Zhao \latin{et~al.}(2024)Zhao, Wang, Yan, Li, Xu, Wuu, Wu, Lai, Lien, and Zhu]{Zhao2024}
Zhao,~M.-J.; Wang,~Y.-T.; Yan,~J.-H.; Li,~H.-C.; Xu,~H.; Wuu,~D.-S.; Wu,~W.-Y.; Lai,~F.-M.; Lien,~S.-Y.; Zhu,~W.-Z. Dielectric properties of hafnium oxide film prepared by HiPIMS at different O$_2$/Ar ratios and their influences on TFT performance. \emph{Journal of Science: Advanced Materials and Devices} \textbf{2024}, \emph{9}, 100722\relax
\mciteBstWouldAddEndPuncttrue
\mciteSetBstMidEndSepPunct{\mcitedefaultmidpunct}
{\mcitedefaultendpunct}{\mcitedefaultseppunct}\relax
\EndOfBibitem
\bibitem[Sharath \latin{et~al.}(2014)Sharath, Bertaud, Kurian, Hildebrandt, Walczyk, Calka, Zaumseil, Sowinska, Walczyk, Gloskovskii, Schroeder, and Alff]{Sharath2014-sg}
Sharath,~S.~U.; Bertaud,~T.; Kurian,~J.; Hildebrandt,~E.; Walczyk,~C.; Calka,~P.; Zaumseil,~P.; Sowinska,~M.; Walczyk,~D.; Gloskovskii,~A.; Schroeder,~T.; Alff,~L. Towards Forming-Free Resistive Switching in Oxygen Engineered HfO$_{2-x}$. \emph{Appl. Phys. Lett.} \textbf{2014}, \emph{104}, 063502\relax
\mciteBstWouldAddEndPuncttrue
\mciteSetBstMidEndSepPunct{\mcitedefaultmidpunct}
{\mcitedefaultendpunct}{\mcitedefaultseppunct}\relax
\EndOfBibitem
\bibitem[Sharath \latin{et~al.}(2014)Sharath, Kurian, Komissinskiy, Hildebrandt, Bertaud, Walczyk, Calka, Schroeder, and Alff]{Sharath2014}
Sharath,~S.~U.; Kurian,~J.; Komissinskiy,~P.; Hildebrandt,~E.; Bertaud,~T.; Walczyk,~C.; Calka,~P.; Schroeder,~T.; Alff,~L. Thickness independent reduced forming voltage in oxygen engineered HfO$_2$ based resistive switching memories. \emph{Applied Physics Letters} \textbf{2014}, \emph{105}, 073505\relax
\mciteBstWouldAddEndPuncttrue
\mciteSetBstMidEndSepPunct{\mcitedefaultmidpunct}
{\mcitedefaultendpunct}{\mcitedefaultseppunct}\relax
\EndOfBibitem
\bibitem[Chen \latin{et~al.}()Chen, Haddad, Wu, Fang, Lan, Avanzino, Pangrle, Buynoski, Rathor, Cai, Tripsas, Bill, Buskirk, and Taguchi]{AnChen}
Chen,~A.; Haddad,~S.; Wu,~Y.-C.; Fang,~T.-N.; Lan,~Z.; Avanzino,~S.; Pangrle,~S.; Buynoski,~M.; Rathor,~M.; Cai,~W.; Tripsas,~N.; Bill,~C.; Buskirk,~M.~V.; Taguchi,~M. Non-Volatile Resistive Switching for Advanced Memory Applications. {IEEE} {International Electron} Devices Meeting, 2005. {IEDM} Technical Digest.\relax
\mciteBstWouldAddEndPunctfalse
\mciteSetBstMidEndSepPunct{\mcitedefaultmidpunct}
{}{\mcitedefaultseppunct}\relax
\EndOfBibitem
\bibitem[Winkler \latin{et~al.}(2022)Winkler, Zintler, Petzold, Piros, Kaiser, Vogel, Nasiou, McKenna, Molina‐Luna, and Alff]{Winkler2022}
Winkler,~R.; Zintler,~A.; Petzold,~S.; Piros,~E.; Kaiser,~N.; Vogel,~T.; Nasiou,~D.; McKenna,~K.~P.; Molina‐Luna,~L.; Alff,~L. Controlling the Formation of Conductive Pathways in Memristive Devices. \emph{Advanced Science} \textbf{2022}, \emph{9}\relax
\mciteBstWouldAddEndPuncttrue
\mciteSetBstMidEndSepPunct{\mcitedefaultmidpunct}
{\mcitedefaultendpunct}{\mcitedefaultseppunct}\relax
\EndOfBibitem
\bibitem[Petzold \latin{et~al.}(2019)Petzold, Zintler, Eilhardt, Piros, Kaiser, Sharath, Vogel, Major, McKenna, Molina‐Luna, and Alff]{Petzold2019}
Petzold,~S.; Zintler,~A.; Eilhardt,~R.; Piros,~E.; Kaiser,~N.; Sharath,~S.~U.; Vogel,~T.; Major,~M.; McKenna,~K.~P.; Molina‐Luna,~L.; Alff,~L. Forming‐Free Grain Boundary Engineered Hafnium Oxide Resistive Random Access Memory Devices. \emph{Advanced Electronic Materials} \textbf{2019}, \emph{5}\relax
\mciteBstWouldAddEndPuncttrue
\mciteSetBstMidEndSepPunct{\mcitedefaultmidpunct}
{\mcitedefaultendpunct}{\mcitedefaultseppunct}\relax
\EndOfBibitem
\bibitem[Jana and Roy~Chaudhuri(2024)Jana, and Roy~Chaudhuri]{Jana2024}
Jana,~B.; Roy~Chaudhuri,~A. Oxygen Vacancy Engineering and Its Impact on Resistive Switching of Oxide Thin Films for Memory and Neuromorphic Applications. \emph{Chips} \textbf{2024}, \emph{3}, 235–257\relax
\mciteBstWouldAddEndPuncttrue
\mciteSetBstMidEndSepPunct{\mcitedefaultmidpunct}
{\mcitedefaultendpunct}{\mcitedefaultseppunct}\relax
\EndOfBibitem
\bibitem[Kumar \latin{et~al.}(2016)Kumar, Wang, Huang, Kumari, Davila, Strachan, Vine, Kilcoyne, Nishi, and Williams]{Kumar2016}
Kumar,~S.; Wang,~Z.; Huang,~X.; Kumari,~N.; Davila,~N.; Strachan,~J.~P.; Vine,~D.; Kilcoyne,~A. L.~D.; Nishi,~Y.; Williams,~R.~S. Conduction Channel Formation and Dissolution Due to Oxygen Thermophoresis/Diffusion in Hafnium Oxide Memristors. \emph{{ACS} Nano} \textbf{2016}, \emph{10}, 11205--11210\relax
\mciteBstWouldAddEndPuncttrue
\mciteSetBstMidEndSepPunct{\mcitedefaultmidpunct}
{\mcitedefaultendpunct}{\mcitedefaultseppunct}\relax
\EndOfBibitem
\bibitem[Rodriguez-Fernandez \latin{et~al.}(2017)Rodriguez-Fernandez, Cagli, Perniola, Suñé, and Miranda]{RodriguezFernandez2017}
Rodriguez-Fernandez,~A.; Cagli,~C.; Perniola,~L.; Suñé,~J.; Miranda,~E. Effect of the voltage ramp rate on the set and reset voltages of ReRAM devices. \emph{Microelectronic Engineering} \textbf{2017}, \emph{178}, 61–65\relax
\mciteBstWouldAddEndPuncttrue
\mciteSetBstMidEndSepPunct{\mcitedefaultmidpunct}
{\mcitedefaultendpunct}{\mcitedefaultseppunct}\relax
\EndOfBibitem
\bibitem[Lorenzi \latin{et~al.}(2015)Lorenzi, Rao, Irrera, Suñé, and Miranda]{Lorenzi2015}
Lorenzi,~P.; Rao,~R.; Irrera,~F.; Suñé,~J.; Miranda,~E. A thorough investigation of the progressive reset dynamics in HfO$_2$-based resistive switching structures. \emph{Applied Physics Letters} \textbf{2015}, \emph{107}\relax
\mciteBstWouldAddEndPuncttrue
\mciteSetBstMidEndSepPunct{\mcitedefaultmidpunct}
{\mcitedefaultendpunct}{\mcitedefaultseppunct}\relax
\EndOfBibitem
\bibitem[Luo \latin{et~al.}(2012)Luo, Lin, Huang, Lee, and Hou]{Luo2012}
Luo,~W.-C.; Lin,~K.-L.; Huang,~J.-J.; Lee,~C.-L.; Hou,~T.-H. Rapid Prediction of RRAM RESET-State Disturb by Ramped Voltage Stress. \emph{IEEE Electron Device Letters} \textbf{2012}, \emph{33}, 597–599\relax
\mciteBstWouldAddEndPuncttrue
\mciteSetBstMidEndSepPunct{\mcitedefaultmidpunct}
{\mcitedefaultendpunct}{\mcitedefaultseppunct}\relax
\EndOfBibitem
\bibitem[Padovani \latin{et~al.}(2024)Padovani, La~Torraca, Strand, Larcher, and Shluger]{Padovani2024}
Padovani,~A.; La~Torraca,~P.; Strand,~J.; Larcher,~L.; Shluger,~A.~L. Dielectric breakdown of oxide films in electronic devices. \emph{Nature Reviews Materials} \textbf{2024}, \emph{9}, 607–627\relax
\mciteBstWouldAddEndPuncttrue
\mciteSetBstMidEndSepPunct{\mcitedefaultmidpunct}
{\mcitedefaultendpunct}{\mcitedefaultseppunct}\relax
\EndOfBibitem
\bibitem[Su \latin{et~al.}(2018)Su, Wang, Zhang, Wang, Sze, Liu, Chen, Chang, Tsai, Chu, Pan, Wu, and Yang]{Su2018}
Su,~Y.-T.; Wang,~M.-C.; Zhang,~S.; Wang,~H.; Sze,~S.~M.; Liu,~H.-W.; Chen,~P.-H.; Chang,~T.-C.; Tsai,~T.-M.; Chu,~T.-J.; Pan,~C.-H.; Wu,~C.-H.; Yang,~C.-C. A Method to Reduce Forming Voltage Without Degrading Device Performance in Hafnium Oxide-Based 1T1R Resistive Random Access Memory. \emph{IEEE Journal of the Electron Devices Society} \textbf{2018}, \emph{6}, 341–345\relax
\mciteBstWouldAddEndPuncttrue
\mciteSetBstMidEndSepPunct{\mcitedefaultmidpunct}
{\mcitedefaultendpunct}{\mcitedefaultseppunct}\relax
\EndOfBibitem
\bibitem[Portner \latin{et~al.}(2024)Portner, Zellweger, Martinelli, B{\'e}gon-Lours, Bragaglia, Weilenmann, Jubin, Falcone, Hermann, Hrynkevych, Stecconi, La~Porta, Drechsler, Olziersky, Offrein, Gerstner, Luisier, and Emboras]{portner2024actor}
Portner,~K.; Zellweger,~T.; Martinelli,~F.; B{\'e}gon-Lours,~L.; Bragaglia,~V.; Weilenmann,~C.; Jubin,~D.; Falcone,~D.; Hermann,~F.; Hrynkevych,~O.; Stecconi,~T.; La~Porta,~A.; Drechsler,~U.; Olziersky,~A.; Offrein,~B.~J.; Gerstner,~W.; Luisier,~M.; Emboras,~A. {Actor-critic networks with analogue memristors mimicking reward-based learning}. \textbf{2024}, PREPRINT (Version 1) available at Research Square\relax
\mciteBstWouldAddEndPuncttrue
\mciteSetBstMidEndSepPunct{\mcitedefaultmidpunct}
{\mcitedefaultendpunct}{\mcitedefaultseppunct}\relax
\EndOfBibitem
\bibitem[Ulyanenkov(2004)]{Ulyanenkov2004}
Ulyanenkov,~A. LEPTOS: a universal software for x-ray reflectivity and diffraction. Advances in Computational Methods for X-Ray and Neutron Optics. 2004; p 1–15\relax
\mciteBstWouldAddEndPuncttrue
\mciteSetBstMidEndSepPunct{\mcitedefaultmidpunct}
{\mcitedefaultendpunct}{\mcitedefaultseppunct}\relax
\EndOfBibitem
\bibitem[Rycroft(2009)]{Rycroft2009}
Rycroft,~C.~H. VORO++: A three-dimensional Voronoi cell library in C++. \emph{Chaos: An Interdisciplinary Journal of Nonlinear Science} \textbf{2009}, \emph{19}, 041111\relax
\mciteBstWouldAddEndPuncttrue
\mciteSetBstMidEndSepPunct{\mcitedefaultmidpunct}
{\mcitedefaultendpunct}{\mcitedefaultseppunct}\relax
\EndOfBibitem
\bibitem[K\"{u}hne \latin{et~al.}(2020)K\"{u}hne, Iannuzzi, Ben, Rybkin, Seewald, Stein, Laino, Khaliullin, Sch\"{u}tt, Schiffmann, Golze, Wilhelm, Chulkov, Bani-Hashemian, Weber, Bor{\v{s}}tnik, Taillefumier, Jakobovits, Lazzaro, Pabst, and \textit{et al.}]{cp2k}
K\"{u}hne,~T.~D.; Iannuzzi,~M.; Ben,~M.~D.; Rybkin,~V.~V.; Seewald,~P.; Stein,~F.; Laino,~T.; Khaliullin,~R.~Z.; Sch\"{u}tt,~O.; Schiffmann,~F.; Golze,~D.; Wilhelm,~J.; Chulkov,~S.; Bani-Hashemian,~M.~H.; Weber,~V.; Bor{\v{s}}tnik,~U.; Taillefumier,~M.; Jakobovits,~A.~S.; Lazzaro,~A.; Pabst,~H.; \textit{et al.} {CP}2K: An Electronic Structure and Molecular Dynamics Software Package - Quickstep: Efficient and Accurate Electronic Structure Calculations. \emph{J. Chem. Phys.} \textbf{2020}, \emph{152}, 194103\relax
\mciteBstWouldAddEndPuncttrue
\mciteSetBstMidEndSepPunct{\mcitedefaultmidpunct}
{\mcitedefaultendpunct}{\mcitedefaultseppunct}\relax
\EndOfBibitem
\bibitem[Perdew \latin{et~al.}(1996)Perdew, Burke, and Ernzerhof]{pbe}
Perdew,~J.~P.; Burke,~K.; Ernzerhof,~M. Generalized Gradient Approximation Made Simple. \emph{Phys. Rev. Lett.} \textbf{1996}, \emph{78}, 3865--3868\relax
\mciteBstWouldAddEndPuncttrue
\mciteSetBstMidEndSepPunct{\mcitedefaultmidpunct}
{\mcitedefaultendpunct}{\mcitedefaultseppunct}\relax
\EndOfBibitem
\bibitem[VandeVondele and Hutter(2007)VandeVondele, and Hutter]{dzvp}
VandeVondele,~J.; Hutter,~J. {Gaussian basis sets for accurate calculations on molecular systems in gas and condensed phases}. \emph{J. Chem. Phys.} \textbf{2007}, \emph{127}, 114105\relax
\mciteBstWouldAddEndPuncttrue
\mciteSetBstMidEndSepPunct{\mcitedefaultmidpunct}
{\mcitedefaultendpunct}{\mcitedefaultseppunct}\relax
\EndOfBibitem
\bibitem[Kindsm\"{u}ller \latin{et~al.}(2018)Kindsm\"{u}ller, Schmitz, Wiemann, Skaja, Wouters, Waser, Schneider, and Dittmann]{Kindsmller2018}
Kindsm\"{u}ller,~A.; Schmitz,~C.; Wiemann,~C.; Skaja,~K.; Wouters,~D.~J.; Waser,~R.; Schneider,~C.~M.; Dittmann,~R. Valence change detection in memristive oxide based heterostructure cells by hard X-ray photoelectron emission spectroscopy. \emph{{APL} Materials} \textbf{2018}, \emph{6}, 046106\relax
\mciteBstWouldAddEndPuncttrue
\mciteSetBstMidEndSepPunct{\mcitedefaultmidpunct}
{\mcitedefaultendpunct}{\mcitedefaultseppunct}\relax
\EndOfBibitem
\bibitem[Ducry \latin{et~al.}(2020)Ducry, Aeschlimann, and Luisier]{Ducry2020}
Ducry,~F.; Aeschlimann,~J.; Luisier,~M. Electro-Thermal Transport in Disordered Nanostructures: a Modeling Perspective. \emph{Nanoscale Adv.} \textbf{2020}, \emph{2}, 2648--2667\relax
\mciteBstWouldAddEndPuncttrue
\mciteSetBstMidEndSepPunct{\mcitedefaultmidpunct}
{\mcitedefaultendpunct}{\mcitedefaultseppunct}\relax
\EndOfBibitem
\bibitem[Shen \latin{et~al.}(2023)Shen, Vaca, Gibson, and Kumar]{Shen2023}
Shen,~W.; Vaca,~D.; Gibson,~G.; Kumar,~S. Temperature Dependent Thermal Properties of Thin Film Hafnium Oxide. 2023 22nd IEEE Intersociety Conference on Thermal and Thermomechanical Phenomena in Electronic Systems (ITherm). 2023; p 1–6\relax
\mciteBstWouldAddEndPuncttrue
\mciteSetBstMidEndSepPunct{\mcitedefaultmidpunct}
{\mcitedefaultendpunct}{\mcitedefaultseppunct}\relax
\EndOfBibitem
\bibitem[Panzer \latin{et~al.}(2009)Panzer, Shandalov, Rowlette, Oshima, Chen, McIntyre, and Goodson]{Panzer2009}
Panzer,~M.; Shandalov,~M.; Rowlette,~J.; Oshima,~Y.; Chen,~Y.~W.; McIntyre,~P.; Goodson,~K. Thermal Properties of Ultrathin Hafnium Oxide Gate Dielectric Films. \emph{IEEE Electron Device Letters} \textbf{2009}, \emph{30}, 1269–1271\relax
\mciteBstWouldAddEndPuncttrue
\mciteSetBstMidEndSepPunct{\mcitedefaultmidpunct}
{\mcitedefaultendpunct}{\mcitedefaultseppunct}\relax
\EndOfBibitem
\bibitem[Scott \latin{et~al.}(2018)Scott, Gaskins, King, and Hopkins]{Scott2018}
Scott,~E.~A.; Gaskins,~J.~T.; King,~S.~W.; Hopkins,~P.~E. Thermal conductivity and thermal boundary resistance of atomic layer deposited high-k dielectric aluminum oxide, hafnium oxide, and titanium oxide thin films on silicon. \emph{APL Materials} \textbf{2018}, \emph{6}\relax
\mciteBstWouldAddEndPuncttrue
\mciteSetBstMidEndSepPunct{\mcitedefaultmidpunct}
{\mcitedefaultendpunct}{\mcitedefaultseppunct}\relax
\EndOfBibitem
\end{mcitethebibliography}

\newpage 

\end{document}